\documentclass[iop,apj,tighten,twocol,twocolappendix,numberedappendix]{emulateapj}
\pdfoutput=1 
\usepackage{apjfonts} 
\usepackage{amsmath,amstext}
\usepackage[breaklinks=true,colorlinks,citecolor=blue,linkcolor=magenta]{hyperref} 
\usepackage[all]{hypcap} 
\usepackage{relsize}
\usepackage{rotating}
\usepackage{longtable}
\usepackage{soul}
\def\red#1 {\textcolor{red}{#1}\ }   
\def\blue#1 {\textcolor{blue}{#1}\ }   

\shorttitle{Obliquity of Exoplanet Systems}
\shortauthors{Mu\~noz \& Perets}

\begin{document}

\title{Statistical Trends in the Obliquity Distribution of Exoplanet Systems}  
\author{Diego J. Mu\~noz$^{1,2,3}$ \& Hagai B. Perets$^{3}$}
\affil{
$^1$Center for Interdisciplinary Exploration and Research in Astrophysics, Physics and Astronomy,
Northwestern University, Evanston, IL 60208, USA\\
$^2$ Steward Observatory, University of Arizona, Tucson, AZ 85721, USA\\
$^3$Physics Department, Technion - Israel Institute of Technology, Haifa, Israel 3200003}
\email[Contact: ]{diego.munoz@northwestern.edu}

\begin{abstract}
Important clues on the formation and evolution of planetary systems can be inferred from the stellar obliquity $\psi$. We study the  distribution of obliquities using the 
California-{\it Kepler} Survey and the TEPCat Catalog of Rossiter-MacLaughlin (RM) measurements,
from which we extract, respectively, 275 and 118 targets. We infer a ``best fit'' obliquity distribution in $\psi$ with a single parameter $\kappa$. Large values of $\kappa$ imply that $\psi$ is distributed narrowly around zero, while small values imply approximate isotropy. Our findings are: (1) The distribution of $\psi$ in {\it Kepler} systems is narrower than found by previous studies and consistent with $\kappa\sim15$ (mean $\langle\psi\rangle\sim19^\circ$ and spread $\psi\sim10^\circ$). (2) The value of $\kappa$ in {\it Kepler} systems does not depend, at a statistically significant level, on planet multiplicity, stellar multiplicity or stellar age; on the other hand,  
metal rich hosts, small planet hosts and long-period planet hosts tend to be more oblique than the general sample (at a $\sim$2.5-$\sigma$ significance level).
(3) Hot Jupiter (HJ) systems with RM measurements are consistent with $\kappa\sim2$, more broadly distributed than the general {\it Kepler} population. (4) A separation of the RM sample into cooler ($T_{\rm eff}\lesssim$6250~K) and hotter ($T_{\rm eff}\gtrsim$6250~K) HJ hosts results in two distinct distributions, $\kappa_{\rm cooler}\sim4$ and $\kappa_{\rm hotter}\sim1$ (4-$\sigma$ significance), both more oblique than the {\it Kepler} sample. We hypothesize that the total mass in planets may be behind the increasing obliquity with metallicity and planet radius, and that the period dependence could be due to primordial disk alignment rather than tidal realignment of stellar spin. 
\end{abstract}

\maketitle

\section{Introduction}
The alignment of planetary orbits with the spin axis of their host star is a fundamental feature of exoplanetary 
architectures; one that points directly to the physical mechanisms behind planetary formation. 
In the Solar System, for example, the Sun's stellar spin is tilted with respect to the ecliptic by only $\sim7^\circ$ \citep{car1863,bec05}. Exoplanetary
systems, on the other hand, may exhibit severe spin-orbit misalignments, including nearly polar orientations \citep[e.g., Kepler-63b,][]{san13}.
A complete, predictive theory of planet formation
must explain the origin of both large and small stellar obliquities, being able to discern whether these are inherited from the protoplanetary disk or if they are a consequence of later dynamical interactions. 

The true, three-dimensional obliquity $\psi$ of a planet-hosting star is not a direct observable. This angle can be expressed as \citep[e.g.][]{win07}
\begin{equation}\label{eq:true_obliquity}
\cos\psi=\cos I_* \cos i_{\rm p} +\sin I_* \sin i_{\rm p} \cos\lambda
\end{equation}
where $i_{\rm p}$ is the planet's orbital inclination, $I_*$ is the stellar line-of-sight (LOS) inclination and $\lambda$  is the  projected obliquity onto the plane of the sky.
Typically, $\lambda$ is measured via the Rossiter--McLaughlin (RM; \citealp{ros24,mcl24}) effect \citep[e.g., see][]{que00,oht05,gim06}, or even estimated via stellar spot variability \citep{nut11},
and can be related to $\psi$ via statistical arguments \citep{fab09}. The stellar LOS inclination $I_*$ is more difficult to measure directly; it can be estimated from 
asteroseimology
\citep{giz03,cam16} or inferred from a combination of projected velocity measurements $V\sin I_*$, stellar radii $R_*$ and rotational periods $P_{\rm rot}$ \citep[e.g.][]{win07,hir14,mor14}. Provided that stellar radii and rotational periods are available for a large number of stars (as is the case of the {\it Kepler} catalog; e.g., \citealp{mcq14}),
the $V\sin I_*$  approach
to stellar obliquity inference is the most cost-effective, since 
$V\sin I_*$ measurements are easier to come about than RM ones: not only do they require less spectral resolution and sensitivity, but also do not
need to be taken during transit \citep[e.g.,][]{gau07}

An ensemble of {\it either} $\lambda$ {\it or} $I_*$ measurements
facilitates statistical tests that can constrain to which degree planetary orbits tend to be aligned/misaligned with the spins 
of their host stars \citep{fab09,mor14,cam16}. Spin-orbit statistics can also provide valuable tools for identifying distinct planet populations or
for distinguishing between planet formation models.
Focusing on the origins of hot Jupiters, \citet{mor11} compared the obliquity outcomes of two said models, the Lidov-Kozai migration of model of hot Jupiters by \citet{fab07} and the planet scattering scenario of
\citet{nag08}, inferring that the scattering model more likely given the observations of projected obliquity.

As the number of obliquity measurements increases further, 
astronomers are able to identify other emerging trends that relate the distribution of obliquity to different
planetary and stellar properties. One such trend is that of 
hot Jupiters tending to have smaller values of $\lambda$ when their host stars have effective temperatures below $T_{\rm eff}\sim 6000$-$6300$K 
(\citealp{win10,alb12,alb13}, see also \citealp{win15}).
This temperature dependence is the most robust obliquity relation in the literature, and it has been reflected not only in $\lambda$, but also in the distribution of photometric modulation amplitudes \citep{maz15},
 which should depend on the orientation of the stellar spin axis \citep[see also][]{li16}.
Another possible trend relates hot Jupiter obliquity to stellar age
\citep{tri11}, and there is also evidence
to suggest that hot Jupiters {\it in general} are more oblique than the general planet population
(\citealp{alb13}, and more recently \citealp{win17}). One intriguing finding, perhaps pointing to the
dynamical evolution of planetary systems, is that stars with
multiple planets may be very closely aligned \citep{san12,alb13} although it is unclear
if the converse is true of single-transiting systems: \citealp{mor14} noted a modest trend
suggestive of higher obliquity in systems with one transiting planet, but very recently, \citet{win17}
reported that the trend has disappeared. 

In the present work, we address the statistical trends mentioned above, and explore additional ones, 
making use of two publicly available catalogs: the California-{\it Kepler} Survey  
\citep{pet17,joh17} and the
TEPCat database \citep{sou11}.

\section{Obliquity Distribution from Bayesian Inference}\label{sec:overview}

\subsection{The Geometry of Stellar Obliquity}
The stellar obliquity $\psi$ is the angle between the
stellar spin vector $\mathbf{S}_*$ and the planetary orbit's angular momentum vector $\mathbf{L}_{\rm p}$.
The three-dimensional orientation of $\mathbf{S}_*$ in space is determined by a polar angle $\theta$ and an azimuthal angle $\phi$.
Observationally, however, instead of $\theta$ and $\phi$, it  is more convenient to work in terms of $I_*$, the angle between $\mathbf{S}_*$ the LOS, 
and  $\beta_*$, the projected angle of $\mathbf{S}_*$  onto the plane of the sky.  These angles are related by:
\begin{subequations}\label{eq:angle_transform}
\begin{align}
\label{eq:angle_transform_a}
\cos I_* =\sin\theta\cos\phi~~,\\
\label{eq:angle_transform_b}
\sin I_* \sin\beta_*=\sin\theta\sin\phi~~,\\
\sin I_* \cos \beta_* =\cos\theta~~.
\end{align}
\end{subequations}
Now, for transiting planets, one can assume that $i_{\rm p}\approx90^\circ$ (cf.~Eq.~\ref{eq:true_obliquity}), i.e., $\mathbf{L}_{\rm p}$ is perpendicular to the LOS;
this assumption allows us to set $\theta=\psi$ and $\beta_*=\lambda$, where $\lambda$
is the projected spin-orbit misalignment angle.

In the following, we describe how a collection of $\lambda$ or $\cos I_*$ measurements can be used to constrain the statistical properties
of the true obliquity $\psi$.

\subsection{The Fisher Distribution for $\psi$ and the Concentration Parameter}
\begin{figure}[t!]
\includegraphics[width=0.47\textwidth]{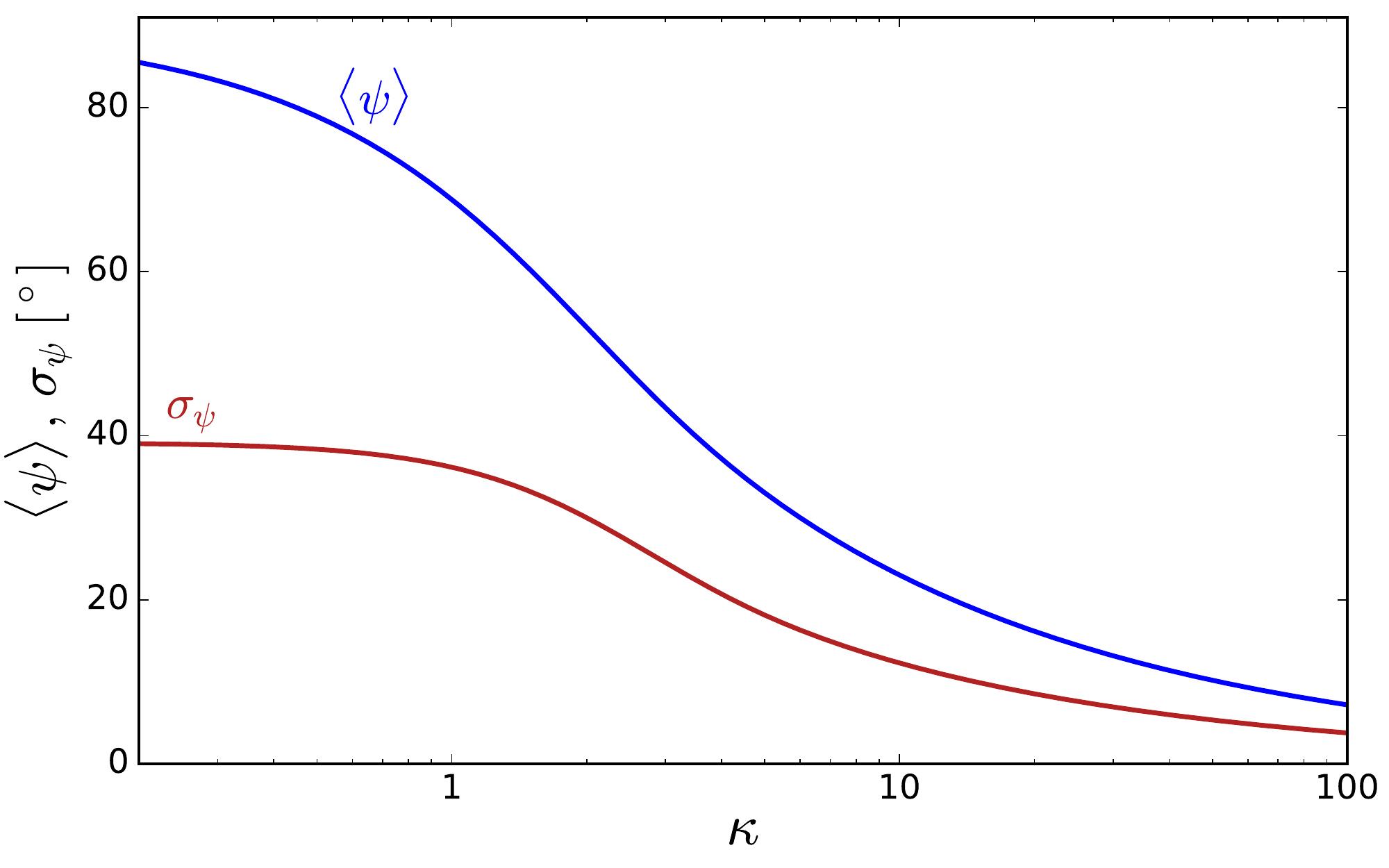}
\caption{{Moments of the Fisher distribution (Eq.~\ref{eq:fisher_dist}) in degrees. The mean obliquity $\langle\psi\rangle$ for a given value of
$\kappa$  is depicted in blue; the spread (standard deviation) of obliquity $\sigma_\psi$ given $\kappa$ is depicted in red. When
$\kappa=0$ (isotropy), $\langle\psi\rangle=90^\circ$ and $\sigma_\psi=({\pi^2/4-2})^{1/2}\approx39.17^\circ$. When $\kappa\gtrsim30$, 
$f_\psi$ approximates a Rayleigh distribution with $\langle\psi\rangle\approx \sqrt{{\pi}/{2}}\kappa^{-1/2}$
and $\sigma_\psi=(4/\pi-1)^{1/2}\sqrt{{\pi}/{2}}\kappa^{-1/2}$.}
\label{fig:fisher_moments}}
\end{figure}

We are interested in finding the distribution of obliquities for a sample of known exo-planetary systems. For this,
it is convenient to have a model function, such as the  Fisher distribution \citep{fis53, fisher93}, which was proposed for exoplanet
obliquities by \citet{fab09} \citep[see also][]{tre12} and has the form
\begin{equation}\label{eq:fisher_dist}
f_\psi(\psi|\kappa)=\frac{\kappa}{2\sinh \kappa}\exp(\kappa\cos\psi)\sin\psi
\end{equation}
where $\kappa$ is often referred to as the ``concentration parameter". The Fisher distribution of Eq.~(\ref{eq:fisher_dist})
does not have closed-form expressions for its moments. The mean $\langle \psi\rangle$ and standard deviation
$\sigma_\psi\equiv\sqrt{\langle \psi^2\rangle-\langle \psi\rangle^2}$ are shown in Fig.~\ref{fig:fisher_moments} as a function of 
$\kappa$. For large $\kappa$, $f_\psi(\psi|\kappa)$ reduces to
the Rayleigh distribution with scale parameter $\sigma\equiv\kappa^{-1/2}$, for which $\langle \psi\rangle=\sqrt{{\pi}/{2}}\sigma$
and $\sigma_\psi=\sqrt{2-\tfrac{\pi}{2}}\sigma$. Thus, the quantity $\kappa^{-1/2}$ 
provides a scale for the mean and spread of the obliquity angle.

Given the actual observable angles, it is convenient to work with the probability distributions of 
$\cos I_*$ or $\lambda$. \citet{mor14} (hereafter MW14) have shown that,
by means of Eq.~(\ref{eq:angle_transform_a}) and standard probability rules, one can derive $f_{\cos I_*}(z=\cos I_*|\kappa)$ in closed form:
\begin{equation}\label{eq:cosi_given_kappa}
f_{\cos I_*}(z|\kappa)=\frac{2\kappa}{\pi\sinh\kappa}\int\limits_z^1\frac{\cosh(\kappa\sqrt{1-y^2})}{\sqrt{1-y^2}}\frac{1}{\sqrt{1-(z/y)^2}}dy
\end{equation}
which is normalized in such a way that $\cos I_*\in[0,1]$.
In this work (see Appendix~\ref{app:appendix_a}), we  also show that a similar derivation can be carried out to compute
the PDF of the projected obliquity $\lambda$ given $\kappa$:
\begin{equation}\label{eq:lambda_given_kappa}
\begin{split}
f_{\lambda}(\lambda |\kappa)&=\frac{\kappa}{2\pi\sinh\kappa}\frac{1}{\cos^2\lambda}
\int\limits_0^{\cos\lambda}\frac{\exp(\kappa{y})}{\sqrt{1-{y}^2}}
\frac{{y}}{\sqrt{1-\left(\tfrac{{y}^2}{1-{y}^2}\right)\tan^2\lambda}}
d{y}
\end{split}
\end{equation}
normalized to be valid in the range $\lambda\in[-\pi,\pi]$. To the extent of our knowledge,
this expression for $f_{\lambda}(\lambda |\kappa)$ has not been presented previously in the
literature.

\subsection{Hierarchical Bayesian Inference}\label{sec:bayesian}
For a given dataset $D$ containing $N$ stars with $N$ measurement posteriors of
some quantity that depends on $\psi$ given $\kappa$, we write
the total likelihood as 
\begin{equation}\label{eq:kappa_likelihood}
\mathcal{L}_{\kappa}=\prod\limits_{n=1}^N\mathcal{L}_{\kappa,n}~~.
\end{equation}
The contribution from each measurement $\mathcal{L}_{\kappa,n}$ to the total likelihood
$\mathcal{L}_{\kappa}$ depends on the stellar quantity being measured. If we have posteriors for $z=\cos I_*$, then we have
\citep[e.g.][]{hog10}
\begin{equation}\label{eq:kappa_deltalike_cosi}
\mathcal{L}_{\kappa,n}(D|\kappa) \propto \int_0^1 dz\, p(z| D_n) \cfrac{f_{\cos I_*}(z|\kappa)}{\pi_{\cos I_*,0}(z)}
\end{equation}
where $\pi_{\cos I_*,0}(z)=1$ for all $z$ is an uninformative prior. Similarly, for another dataset
$\widetilde{D}$ from which $N$ posteriors on $\lambda$ can be obtained, we have
\begin{equation}\label{eq:kappa_deltalike_lambda}
\mathcal{L}_{\kappa,n}(\widetilde{D}|\kappa) \propto \int_{-\pi}^{\pi } d\lambda\, p(\lambda | \widetilde{D}_n) 
\cfrac{f_{\lambda}(\lambda |\kappa)}{\pi_{\lambda,0}(z)}
\end{equation}
where the prior $\pi_{\lambda,0}(z)=1/(2\pi)$ is also uninformative and thus only amounts to a normalization constant.
These two likelihoods (Eqs.~\ref{eq:kappa_deltalike_cosi} and~\ref{eq:kappa_deltalike_lambda}) can be computed by direct numerical
integration or by the method of $K$-samples (see Eq.~\ref{eq:delta_likelihood_ksamp}). 

The prior PDF of the meta-parameter $\kappa$ that must multiply the total likelihood in 
Eq~(\ref{eq:kappa_likelihood}) is usually assumed to take the form \citep{fab09}:
\begin{equation}\label{eq:prior_pdf}
\pi_\kappa(\kappa) \propto {(1+\kappa^2)^{-3/4}}~~. 
\end{equation}
{This prior is chosen to be well behaved when $\kappa=0$ and to become uniform
in the Rayleigh scale parameter $\sigma=\kappa^{-1/2}$ as $\kappa\rightarrow\infty$. This prior may underestimate
$\kappa$ if it is large ($\gg 10$), and a flat prior might pick up the signal of a large concentration parameter if the dataset is small} \citep{cam16}.
{An alternative is to device a prior that, for large $\kappa$, the prior becomes logarithmic uninformative
in $\sigma$ (i.e., a Jeffreys prior on a scale parameter) rather than uniform. This can be accomplished with a prior in $\kappa$ of the form}
\addtocounter{equation}{-1}
\begin{subequations}
\begin{align}
\addtocounter{equation}{+1}
\label{eq:prior_pdf_alt}
\pi'_\kappa \propto {(1+\kappa^2)^{-1/2}}~~,
\end{align}
\end{subequations}
{which satisfies $\pi_\sigma\propto1/\sigma$ for $\kappa\gg1$.
For the sake of consistency with previous studies, we will 
employ
the prior function of {Eq.~(\ref{eq:prior_pdf})} unless stated otherwise. As it turns out, the dataset is large enough that the
inference on $\kappa$ is weakly sensitive to the choice of either prior.}
Further details  of the Bayesian computation are provided in Appendix~\ref{app:background}.

\vspace{0.2in}

We note that, in principle, inference using $\cos I_*$ and $\lambda$ simultaneously could be done by writing a two-dimensional
integral for $\mathcal{L}_{\kappa,n}$ in place of Eqs.~(\ref{eq:kappa_deltalike_cosi})-(\ref{eq:kappa_deltalike_lambda})
and using a joint probability
$p (z,\lambda|\kappa)=f_{\cos I_*}(z|\kappa)\times f_{\lambda}(\lambda |\kappa)$, where measurement posteriors for
both $\cos I_*$ and $\lambda$ can be obtained for every object.
 Unfortunately, the number of {\it Kepler} systems
for which both $\cos I_*$ and $\lambda$ has been derived/measured is small (we identify 5 such objects in  Section~\ref{sec:TEPCAT}). Thus, for the remainder of the paper, we will compute
$p(\kappa|D)$ for the different datasets independently, acknowledging that each data set might
sample different population planetary system and thus differences in the inferred values of $\kappa$
are to be expected.

\section{Obliquity Distribution from Observations}

\subsection{The California {\it Kepler} Survey}\label{sec:CKS} 
Recently, \citet{win17} used the California-{\it Kepler} Survey \citep[CKS;][]{pet17,joh17}
to study the statistical properties of line-of-sight inclination $I_*$ and projected rotational
velocity $V\sin I_*$ of numerous {\it Kepler} planet hosts. The extensive analysis of
\citet{win17} did not include inference on the concentration parameter $\kappa$ (Eq.~\ref{eq:fisher_dist}), thus,
the analysis presented below is highly complementary
to their work.

We are interested in objects for which $V\sin I_*$, the stellar radius $R_*$ and
the rotational period $R_*$ are known. The CKS catalog contains 1305 KOIs
with $V\sin I_*$ and $R_*$ measurements, of which 773 have a ``confirmed planet'' disposition \citep{pet17}. To assign rotational periods, we collect $P_{\rm rot}$ measurements
from two main catalogs: \citet{maz15}, which obtained via auto-correlation function analysis, and
\citet{ang18}, who introduced a novel Bayesian parameter estimation of $P_{\rm rot}$
as part of a parametric model consisting of a quasi-periodic
Gaussian random process (QPGP). We are able to find additional periods in the literature, recovering
measurements from \citet{bon12}, \citet{mcq13a}, \citet{mcq14}, \citet{hir14}, \citet{gar14}, \citet{paz15} and \citet{buz16}, all of which were obtained using somewhat different, but still deeply
related methods.  Most of these period identification techniques are based on Fourier analysis
of time series \citep[see discussion in][]{aig15}, with one departure being
the Morlet wavelet method of \citet{gar14},  which is still inherently a spectral
analysis method. We refer to all these strategies collectively as ``spectral analysis'' (SA) methods and group the corresponding periods along with the \citet{maz15} catalog, leaving the \citet{ang18} catalog
as the one truly distinct approach to period identification. Of the 773 entries in the CKS sample,
we are able to assign SA periods for 734, and QPGP periods for 645; 614 targets
have both SA and QPGP periods.
Following \citet{win17}, we proceed with our analysis only using targets with ``reliable'' periods,
meaning those for which SA and QPGP estimates coincide. Our period selection method is analogous although slightly more permissive that the one used by  \citet{win17}. First, we only consider
period measurements with a signal-to-noise ratio of 3 and larger. Second, we consider that
the two period estimates match if they are within 30$\%$ of the identity, {\it provided} {that the uncertainties
overlap (we use 3-$\sigma$ error bars)}. Period filtering removes 492 targets, and 
we are left with 291. Possibly blended sources within the {\it Kepler} photometric aperture 
\citep[$\sim4''$][]{mull15,fur17} means that we cannot attribute
the rotational period to the planet host; therefore, we remove 34 additional targets for which a nearby companion was detected by \citet{fur17} within 3 mag of that of the target KIC star \citep{win17}.
We are left with a database of 257 targets.

\begin{figure}[t!]
\includegraphics[width=0.41\textwidth]{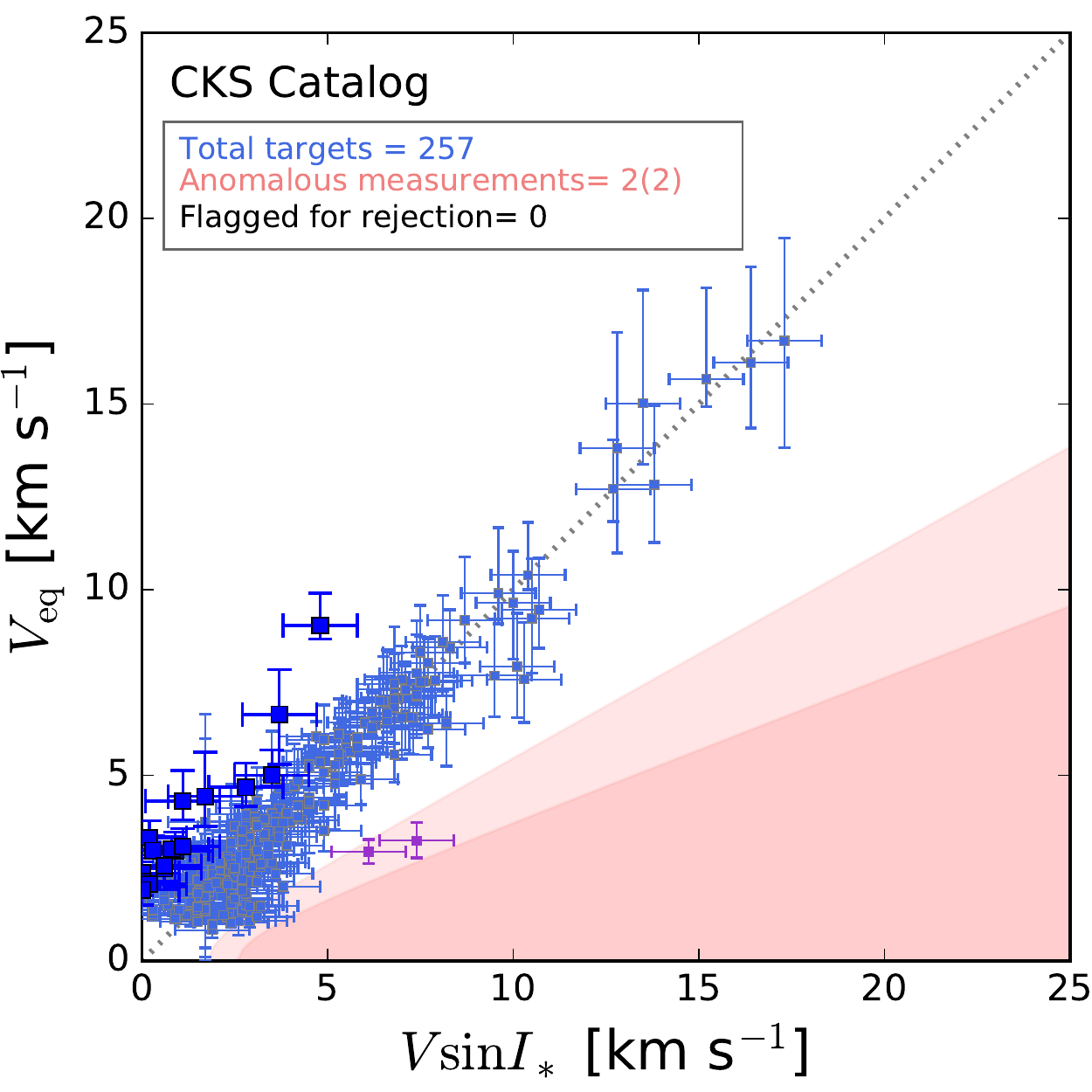}
\vspace{-0.05in}
\includegraphics[width=0.41\textwidth]{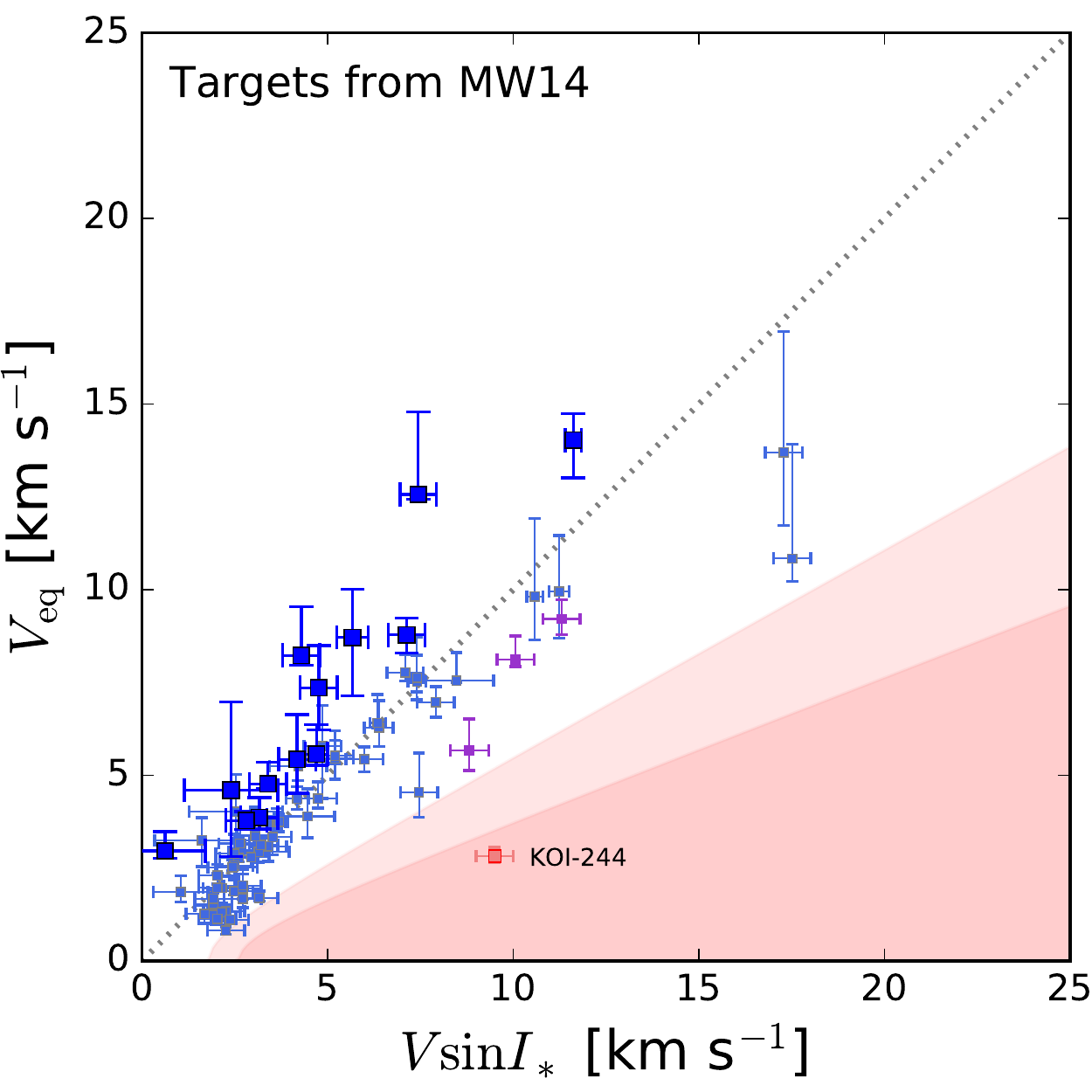}
\caption{Top panel: Inferred equatorial rotational velocities $V_{\rm eq}$ and their $68\%$ confidence
intervals (see text) versus $V\sin I_*$ measurements and their uncertainties for 257 KOIs in the CKS
\citep{pet17}. Data points cluster around the identity, as it is to be expected from low obliquity
systems. Seventeen systems with statistically significant misalignment ($I_*(95\%)\leq86^\circ$) are depicted
by large, dark blue squares, the rest of the objects are depicted by small, light blue squares.
Two data points appear to be anomalous (in purple), as they lie below the identity at a distance
of more than 3-$\sigma$ (but less than 5-$\sigma$) according to their own uncertainties. The two shaded areas are defined
by the curves $V_{\rm eq}=V\sin I_*/1.8$ (upper) and $V_{\rm eq}=V\sin I_*/2.6$ (lower) and serve as 
target rejection boundaries. We define a criterion
for target removal in which all data points for which the $95\%$ upper limit in $V_{\rm eq}$ lies
below the first curve are rejected. In the 257-target dataset  we generated from the CKS catalog, no
object is flat-out rejected.
Bottom panel: Same as above, but for the 70-target dataset of MW14. In that work, 12 targets
are identified as misaligned with respect to the planetary orbits (i.e., confidently different from edge-on),
3 are anomalous, and 1 (KOI-244) is regarded as unphysical and thus rejected.
\vspace{-0.1in}
\label{fig:veqvsvsini_cks}}
\end{figure}

\vspace{-0.1in}
\subsubsection{Inclination Posterior for Each Target}
For each of the 257 CKS targets, we derive a posterior PDF of the line-of-sight 
inclination. In principle, the value of $I_*$ can be obtained from the straightforward operation 
\begin{equation}
\sin I_* = \frac{(V\sin I_*)}{V_{\rm eq}}= \frac{(V\sin I_*)}{2\pi R_*/P_{\rm rot}}
\end{equation}
\citep[e.g.][]{bor84,doy84,sod85,win07,alb11}. However, this approach has been long recognized as nontrivial because the often large uncertainties in the different quantities
involved.


As described in detail by MW14, if one has {PDFs for the projected rotational velocity $V\sin I_*$ and 
the equatorial rotational velocity $V_{\rm eq}$} -- $p_{V_s}$ and $p_{V_{\rm eq}}$
respectively-- the posterior
PDF of $\sin I_*$ given a prior $\pi_{\sin I_*}(z)=z/\sqrt{1-z^2}$ 
(i.e., uniform in $\cos I_*$) is given by\footnote{
To obtain the likelihood of $\sin I_*$,  $\mathcal{L}(\sin I_*)=p(D|\sin I_*)$, one
uses the transformation rule for the quotient of two random variables 
$Z\equiv X/Y$ where the PDFs $f_X(x)$ and $f_Y(y)$ are known:
\begin{equation}
\nonumber f_Z(z) = \int_{-\infty}^{+\infty}|y|f_X(zy)f_Y(y)dy~~.
\end{equation}
}, 
\begin{equation}\label{eq:sin_pdf}
\begin{split}
p(\sin I_* | D)\propto & ~~\mathcal{L}(\sin I_*) \pi_{\sin I_*}(\sin I_*)\\
=& ~~\frac{\sin I_*}{\sqrt{1-\sin^2I_*}}\int_{0}^\infty v' p_{V_s}(v'\sin I_*)p_{V_{\rm eq}}(v')dv'~~.
\end{split}
\end{equation}
It is convenient to work in terms of $\cos I_*$. Since 
$p(\cos I_* | D)=p(\sin I_* | D)|\tfrac{d\sin I_*}{d\cos I_*}|$, the
desired posterior is
\begin{equation}\label{eq:cosi_pdf}
\begin{split}
p(\cos I_* | D)= & p(\sin I_* | D) \frac{\cos I_*}{\sqrt{1-\cos^2I_*}}\\
\propto & ~~\int_{0}^\infty v'p_{V_s}(v'\sqrt{1-\cos^2I_*})p_{V_{\rm eq}}(v')dv'~~.
\end{split}
\end{equation}
MW14 assume $p_{V_s}(\cdot)$ is a Gaussian, but for $p_{V_{\rm eq}}(\cdot)$, they derive an empirical PDF from the Monte Carlo sampling
of $P_{\rm rot}$ and $R_*$  -both assumed to be (double-sided) Gaussians-- after incorporating the effects of differential rotation and the stochasticity of stellar spots.  From the PDF $p_{V_{\rm eq}}$, we derive a most likely value $V_{\rm eq}$ and a confidence interval and plot it against 
the measured value of $V\sin I_*$; this is shown in Fig.~\ref{fig:veqvsvsini_cks} (top panel).
All targets above the  $V_{\rm eq}=V\sin I_*$ line (i.e., $\sin I_*<1$) 
are not edge-on (and potentially misaligned with respect to
the planetary orbits). Targets for which $\sin I_*<1$ with some statistical certainty are highlighted.
Targets below the $V_{\rm eq}=V\sin I_*$ line (i.e., $\sin I_*>1$)
 are unphysical if the uncertainties cannot account for such a measurement (see figure
caption for further details). Fig.~\ref{fig:veqvsvsini_cks} (bottom panel) depicts the same
comparison of velocities, but for the MW14 database. In our final 257-target database, two targets
-- KOIs 94 and 1848 (Kepler-89 and Kepler-978 respectively) -- appear to be marginally unphysical, 
i.e., they 
are below the identity line at a distance greater than 3-$\sigma$ but smaller than 5-$\sigma$.
We deem these two targets ``anomalous'', but we do not remove them. Note that in the MW14 database, four targets are anomalous, but only one (KOI 244) is removed from their analysis.
If not filtered out,  anomalous targets will still have well behaved 
$\cos I_*$ PDFs (Eq.~\ref{eq:cosi_pdf}) that strongly peak at $\cos I_*=0$ 
(see Fig.~\ref{fig:cosipdf_cks} below), and favor lower values
of $\kappa$ when entering Eq.~(\ref{eq:kappa_deltalike_cosi}).

The 257 PDFs of $\cos I_*$ are shown in Fig.~\ref{fig:cosipdf_cks}. Of all these targets,
only 17 (highlighted curves) have $I_*\neq90^\circ$ to a $95\%$ confidence level (e.g., see MW14).
This fraction ($6.6\%$) is lower than in the 70-target sample of MW14, which concluded that the number
of misaligned (not edge-on) stars was 12 (or $17\%$). We thus expect the overall population
to have lower mean obliquities, or a higher value of $\kappa$ in the distribution
of Eq.~(\ref{eq:fisher_dist}), than originally reported by MW14.

\begin{figure}[t!]
\includegraphics[width=0.47\textwidth]{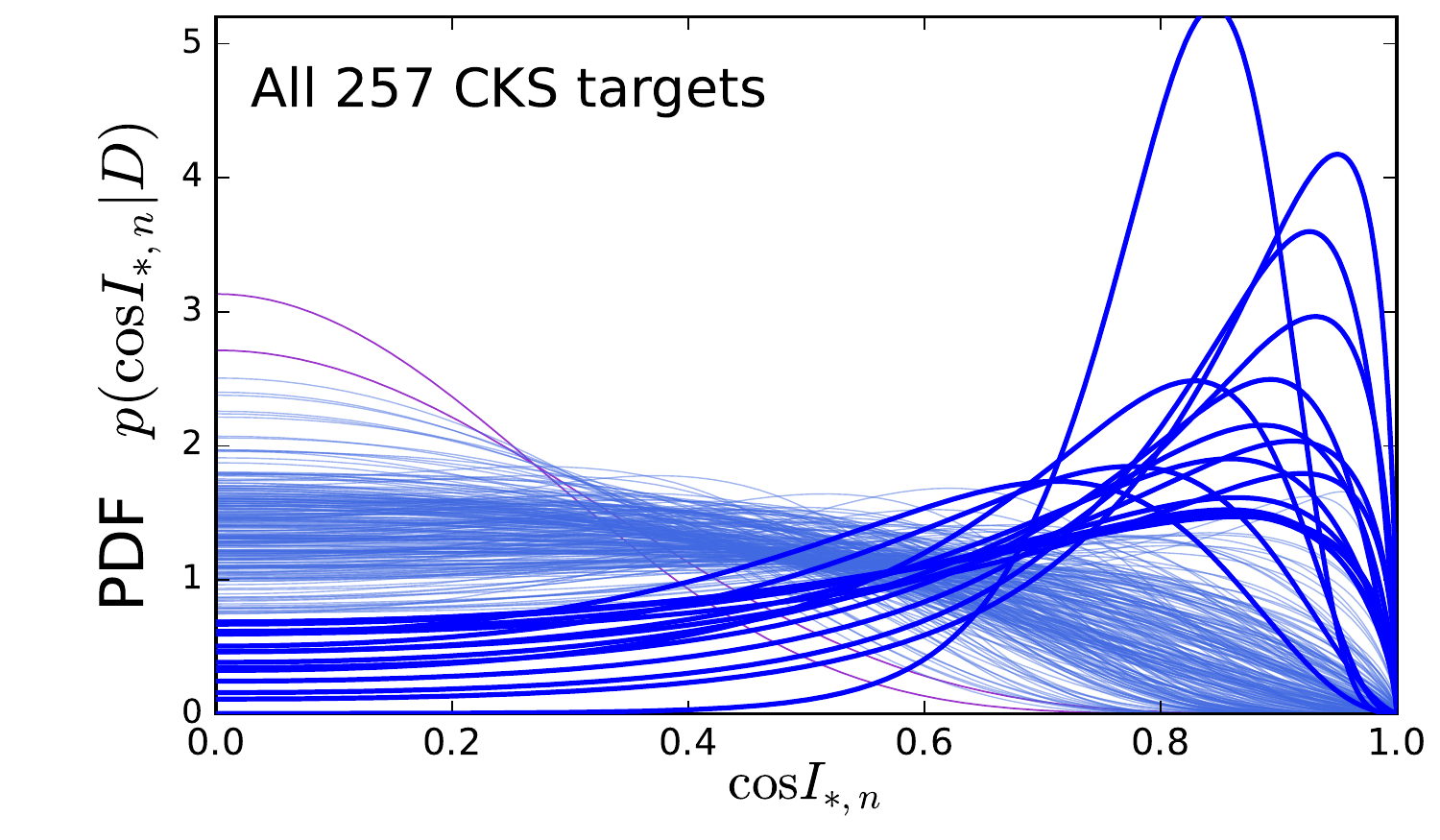}
\caption{Posterior PDFs of $\cos I_*$ (Eq.\ref{eq:cosi_pdf})
for the 257 targets in the dataset we compiled from the CKS catalog. The 17 targets that have
LOS inclinations significantly different from edge-on are highlighted. The two targets
that are partially anomalous (KOIs 94 and 1848; purple data points in Fig.~\ref{fig:veqvsvsini_cks}) have PDFs that are strongly concentrated toward $\cos I_*=0$. Statistically significant...
\label{fig:cosipdf_cks}}
\end{figure}

\subsubsection{Obliquity Distribution}
\label{sec:obliquity_distribution}
\begin{figure}[t!]
\includegraphics[width=0.47\textwidth]{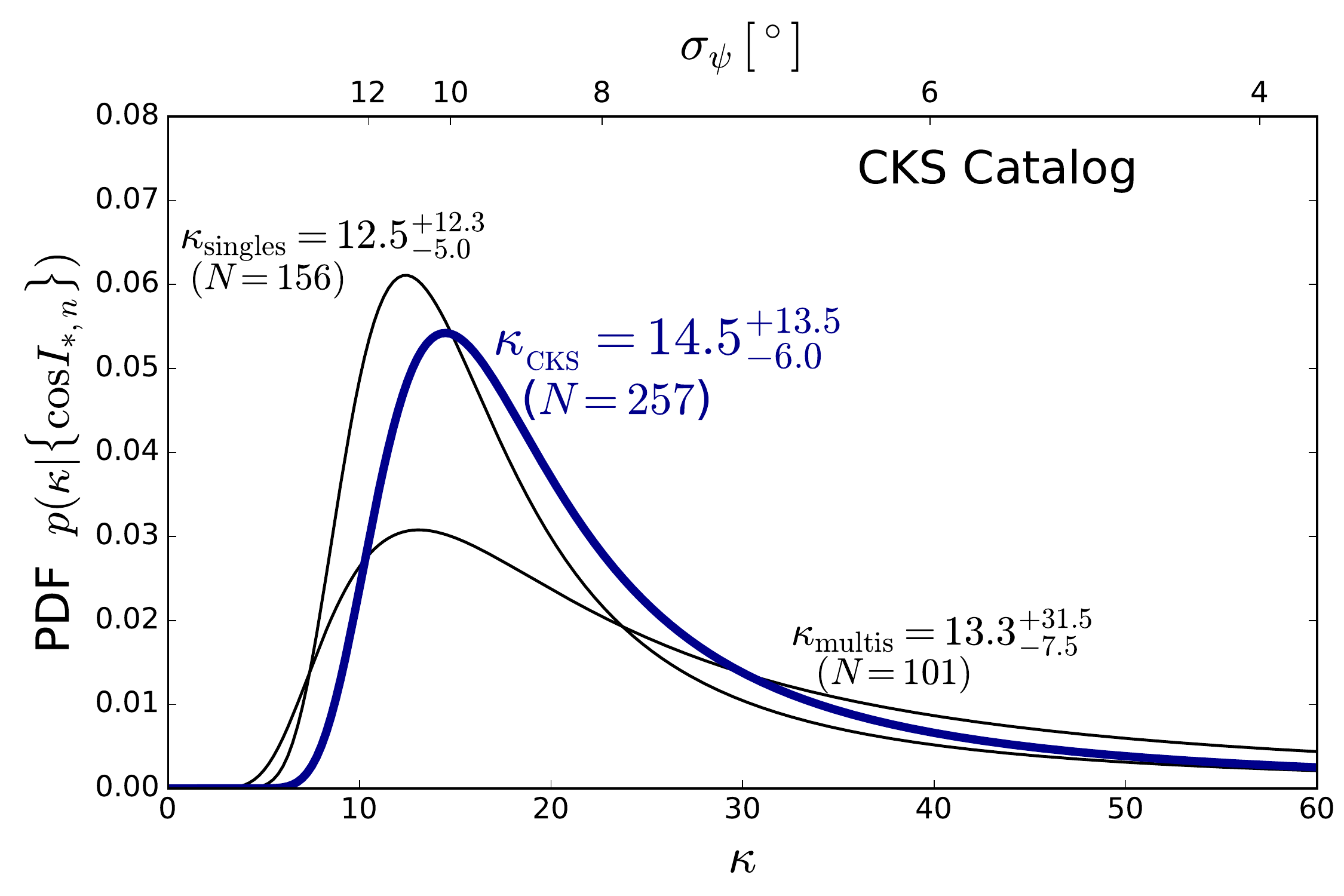}
\vspace{-0.05in}
\caption{Posterior PDF of the $\kappa$ parameter (Eq.~\ref{eq:kappa_likelihood} multiplied by Eq.~\ref{eq:prior_pdf} using Eq.~\ref{eq:kappa_deltalike_cosi}) in dark blue. The distribution is summarized
by the representative value of 14.4 and a $68\%$-probability interval  $[8.8, 28]$, i.e.,
$\kappa_{\text{\tiny CKS}}=14.5^{+13.5}_{-6}$.
 In addition, we present the $\kappa$ posterior
from two subsets: systems with one transiting planet and systems with multiple transiting planets.
The two subsets produce virtually indistinguishable concentration parameter PDFs 
(thin black curves), suggesting
that the underlying obliquity does not depend on planet multiplicity.
\label{fig:kappa_posterior_cks}}
\vspace{-0.05in}
\end{figure}

With the 257 $\cos I_*$ posteriors depicted in Fig.~\ref{fig:cosipdf_cks}, we can
carry out the hierarchical Bayesian inference method summarized in Section~\ref{sec:overview}.
The likelihood of the data given $\kappa$ is computed from Eqs.~(\ref{eq:kappa_likelihood})
and~(\ref{eq:kappa_deltalike_cosi}), and the prior used is given by Eq.~(\ref{eq:prior_pdf}).
The resulting posterior PDF of $\kappa$ is shown in Fig.~\ref{fig:kappa_posterior_cks}, with a
(weighted) maximum a posteriori\footnote{
As the maximum a posteriori (MAP) or mode of a posterior PDF $f_X(x)$ is often an inconvenient ``best fit'' value of a parameter $X$, we introduce a ``weighted MAP" estimator, which
consists of the mean value of $f_X(x)$ only for the interval where $\{f_X(x) >0.98\max(f)\}$. Thus,
the weighted MAP represents a compromise of sorts between the MAP and the median.
} of 14.5  and a shortest $68\%$-probability interval of $\kappa\in[8.5, 28.0]$; thus, 
we write $\kappa_{\text{\tiny CKS}}=14.5^{+13.5}_{-6}$. This concentration parameter is significantly
larger than the value reported by MW14, which we re-derived to be
$\kappa_{\text{\tiny MW}}=6.2^{+1.8}_{-1.6}$. A value of 
$\kappa=15$ in Eq.~(\ref{eq:fisher_dist}) corresponds to a mean obliquity of
$\langle\psi\rangle=19^\circ$ and a standard deviation of  $\sigma_\psi=10^\circ$,
while $\kappa=6$ results in $\langle\psi\rangle=30^\circ$ and $\sigma_\psi=16^\circ$.
This ``flattening" of the obliquity distribution might be a natural consequence of
a larger number of systems with smaller planets being added to the list,
revealing that the vast majority of systems have low obliquities. If we had removed KOIs 95 and 1848
(which are marginally unphysical and favor low obliquities; purple data points in Figs.\ref{fig:veqvsvsini_cks} and~\ref{fig:cosipdf_cks}) we would have obtained $\kappa_{\text{\tiny CKS}}=13.8^{+12.2}_{-3.8}$, and thus their influence on $\kappa$ is negligible.
{We have also checked the influence of the prior function $\pi_\kappa$ of {Eq.~(\ref{eq:prior_pdf})} versus the alternative
prior $\pi_\kappa'$ of {Eq.~(\ref{eq:prior_pdf_alt}}). If $\pi_\kappa'$ is used, we find that the posterior $p(\kappa|D)$ produces 
$\kappa_{\text{\tiny CKS}}=15.4^{+23.8}_{-6.6}$, as expected from $\pi'_\kappa$ being 
a more slowly declining function of $\kappa$, but a minor
change considering the uncertainties. Up to this point, we can conclude that the CKS test is more spin-orbit aligned than
the dataset used by MW14 and that the sample is consistent with obliquities in the range $\psi=(19\pm10)^\circ$, provided that
the Fisher distribution is an adequate model (see below).}

A tentative trend discovered by MW14, which now seems to have disappeared \citep{win17},
is that of $\kappa$ having a dependence on planet multiplicity. In Fig.~\ref{fig:kappa_posterior_cks},
we show the separation of the CKS
dataset into a subset containing ``singles'' (systems with one transiting planet) and one containing ``multis'' (systems with multiple transiting planets). Indeed, the multiplicity trend no longer appears
to be real, as we find $\kappa^{\text{\tiny CKS}}_{\rm singles}=12.2^{+12.3}_{-5}$ and
$\kappa^{\text{\tiny CKS}}_{\rm multis}=13.3^{+31.5}_{-7.5}$, i.e., the two posterior PDFs are
indistinguishable. 

\begin{figure}[t!]
\includegraphics[width=0.47\textwidth]{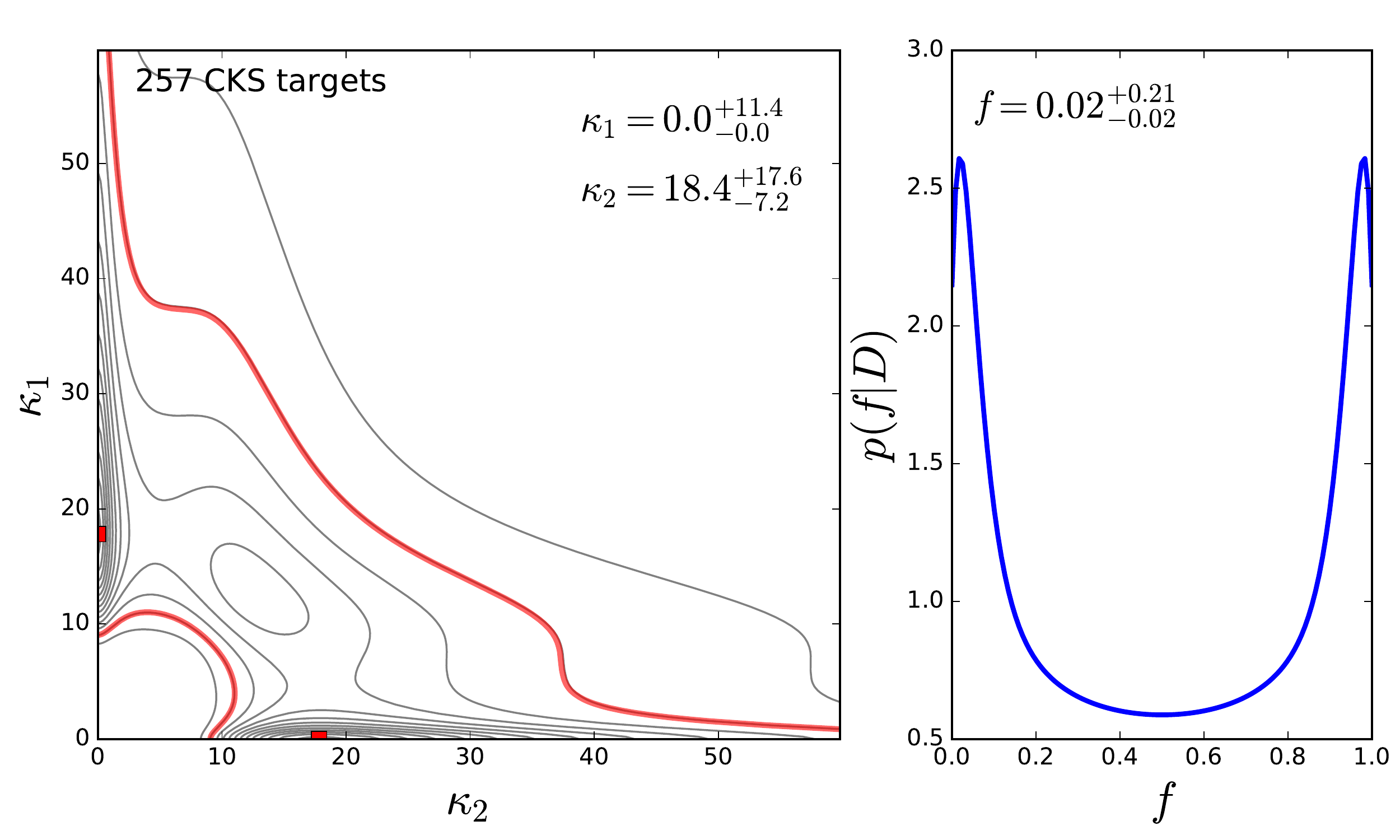}
\vspace{-0.05in}
\caption{{Posterior PDF for the two-Fisher mixture model of three parameters $\kappa_1$, $\kappa_2$ and $f$ (Eq.~{\ref{eq:joint_posterior}}).
Left panel: joint posterior PDF of $\kappa_1$ and $\kappa_2$ after marginalizing over $f$. Right panel: the posterior of $f$
after marginalizing over $\kappa_1$ and $\kappa_2$.
The $\kappa_1$-$\kappa_2$ mirroring symmetry is expected from the model, as is the symmetry of $f$ around 0.5. For simplicity,
we report on the posterior results assuming $\kappa_1<\kappa_2$, which correspond to $f$  peaking close to zero. }
\label{fig:joint_posterior}}
\vspace{-0.05in}
\end{figure}

\paragraph{Model Selection} 
{Despite its useful functional form, the Fisher distribution {(Eq.~\ref{eq:fisher_dist})} is not fully justified on physical grounds
and, in principle, other obliquity distributions could fit the data better.} \citet{fab09} {considered alternative models, such as a mixture of
Fisher distributions or a mixture of an isotropic distribution and a spin-orbit aligned one.  A mixture model
of two Fisher distributions (two concentration parameters $\kappa_1$ and $\kappa_2$ with relative weights $f$ and $1{-}f$)
has a joint posterior PDF given by}
\begin{equation}
\begin{split}\label{eq:joint_posterior}
p(\kappa_1,\kappa_2,f| D)&\propto p(D| \kappa_1,\kappa_2,f) \times \pi(\kappa_1,\kappa_2,f)
\\
&\propto \left[\prod\limits_{n=1}^N{\mathcal{L}}_{\boldsymbol{\alpha},n}(D_n|\kappa_1,\kappa_2,f)\right] \pi(\kappa_1)\pi(\kappa_2)\pi(f)
\end{split}
\end{equation}
{where the contribution of the $n$-th measurement to the total likelihood  is}
\begin{equation}\label{eq:joint_likelihood}
{\mathcal{L}}_{\boldsymbol{\alpha},n}(D_n|\kappa_1,\kappa_2,f)=f{\mathcal{L}}_{\kappa,n}(D_n|\kappa_1) + (1-f){\mathcal{L}}_{\kappa,n}(D_n|\kappa_2)
\end{equation}
{where ${\mathcal{L}}_{\kappa,n}(D_n|\kappa_1)$ is given by {Eq.~(\ref{eq:kappa_deltalike_cosi})}. In {Eq.~(\ref{eq:joint_posterior})} we have also
assumed that the three-parameter prior $p(\boldsymbol{\alpha})=p(\kappa_1,\kappa_2,f)$ is separable.  Figure~{\ref{fig:joint_posterior}} shows
the joint posterior of $\kappa_1$ and $\kappa_2$ and the marginalized posterior of $f$.
The marginalized ``best fit'' values are $\kappa_1=0^{+11.4}_{-0}$,  $\kappa_2=18.4^{+17.6}_{-7.2}$ and $f=0.02^{+0.21}_{-0.02}$. This
roughly states that the data areconsistent with a small fraction of the population (a few percent, consistent with 17 of of 257 targets being oblique; see
Fig.~{\ref{fig:cosipdf_cks}}) being drawn from a high-obliquity distribution and a large
fraction of the population being drawn from a low-obliquity distribution. }

{To compare this new model (which we call $\mathcal{M}_1$) with the previous, simpler one ($\mathcal{M}_0$), we need to compute}
\begin{equation}
p(D|\mathcal{M})=\int p(D| \kappa_1,\kappa_2,f)\pi(\kappa_1,\kappa_2,f)d\kappa_1d\kappa_2df
\end{equation}
{(sometimes called  the ``Bayesian evidence'') and then calculate the posterior odds ratio:}
\begin{equation}
\begin{split}
\frac{p(\mathcal{M}_0| {D})}{p(\mathcal{M}_1| {D})} &= \underbrace{\frac{p(D|\mathcal{M}_0)}{p(D|\mathcal{M}_1)}}
 \frac{p(\mathcal{M}_0)}{p(\mathcal{M}_1)}\\
 & ~~~~~~\equiv \mathcal{K}
 \end{split}
\end{equation}
{where $\mathcal{K}$ is called the ``Bayes factor'' or "evidence ratio". If we assume that $p(\mathcal{M}_0)=p(\mathcal{M}_1)=1/2$, then the model
choice is given by $\mathcal{K}$. We find that  $\mathcal{K}=2.25$, and thus, the  data supports the null (simplest) model  $\mathcal{M}_0$
in favor of the alternative model $\mathcal{M}_1$ (although not decisively, see} {\citealp{jeffreys61}}, appendix B).

{Alternatively, since  one of the concentration parameters in $\mathcal{M}_1$ is consistent with zero, we can take $\kappa_1=0$ and
create another mixture model ($\mathcal{M}_2$) consistent of an isotropic distribution and a Fisher distribution} \citep[e.g., see][]{cam16}. {The likelihood
function of the $n$-th target becomes}
\begin{equation}\label{eq:joint_likelihood_other}
{\mathcal{L}}_{\boldsymbol{\alpha},n}(D_n|\kappa, f)=f + (1-f){\mathcal{L}}_{\kappa,n}(D_n|\kappa)
\end{equation}
{The posterior distribution of this simpler model (of only two parameters) produces $\kappa=18.1^{+17.4}_{-6.9}$ and $f=0.02^{+0.04}_{-0.01}$.
The evidence ratio, in this case, is $\mathcal{K}=p(D|\mathcal{M}_0)/p(D|\mathcal{M}_2)=12$. Thus, the null model $\mathcal{M}_0$
is substantially favored by the data.}

\vspace{0.1in}
In what follows, we explore the obliquity properties of different subsets -- say $D_A$ and $D_B$ where $\cup \{D_i\}= D$ -- within the CKS catalog. Our aim is to  assess
whether a given subset is 	``more oblique'' or ``less oblique'' than its complement. Although we derive different
values/posteriors of $\kappa_A$ and $\kappa_B$, our main goal is not to analyze the physical meaning of each concentration
parameter, but to measure the statistical significance of the differences encountered. 

\subsubsection{Obliquity Trends: Stellar Properties}
\label{sec:stellar_properties} 
The larger size of the CKS set with respect to previously published catalogs allows us to explore
changes in $\kappa$ as a function of different physical variables.  In the following, we focus on
the properties of the stellar host.

\paragraph{Effective Temperature} One intriguing stellar property
that appears to affect the obliquity of planetary systems is the stellar effective temperature
\citep{sch10,win10,alb12,alb13,maz15,win17}. This transition appears to coincide with
the so-called ``Kraft break'' \citep{str30,sch62,kra67}, identified as
a sharp increase in the measured projected velocities in the field at around $T_{\rm eff}=$6200~K. This
break is attributed to the transition that takes place when the width of the convective envelope of a Solar type star
vanishes, giving rise to radiative envelopes at higher temperatures.
We explore the temperature dependence in our compiled list of 257 {\it Kepler} targets. Placing
a cut at 6200~K, we divide the dataset into ``hotter'' systems ($T_{\rm eff}\geq$6200~K)
and ``cooler systems'' ($T_{\rm eff}<$6200~K), with 20 and 237 targets respectively. We find no statistical difference: we derive $\kappa^{\text{\tiny CKS}}_{\rm hotter}=15.2^{+46}_{-10.4}$ and
$\kappa^{\text{\tiny CKS}}_{\rm cooler}=12.2^{+8.6}_{-4.2}$. However, this inference is largely due
to the small number on KOIs with effective temperatures above 6200~K. This lack of detected planets around ``hotter'' stars is at least in part due to a selection effect, since those hot stars for which
planetary transits are detected tend to be the least variables ones, in turn preventing the measurement
of their rotational periods from photometric variability \citep{maz15,win17}. We return to this
subject in Section~\ref{sec:TEPCAT} below.

\begin{figure}[h!]
\includegraphics[width=0.47\textwidth]{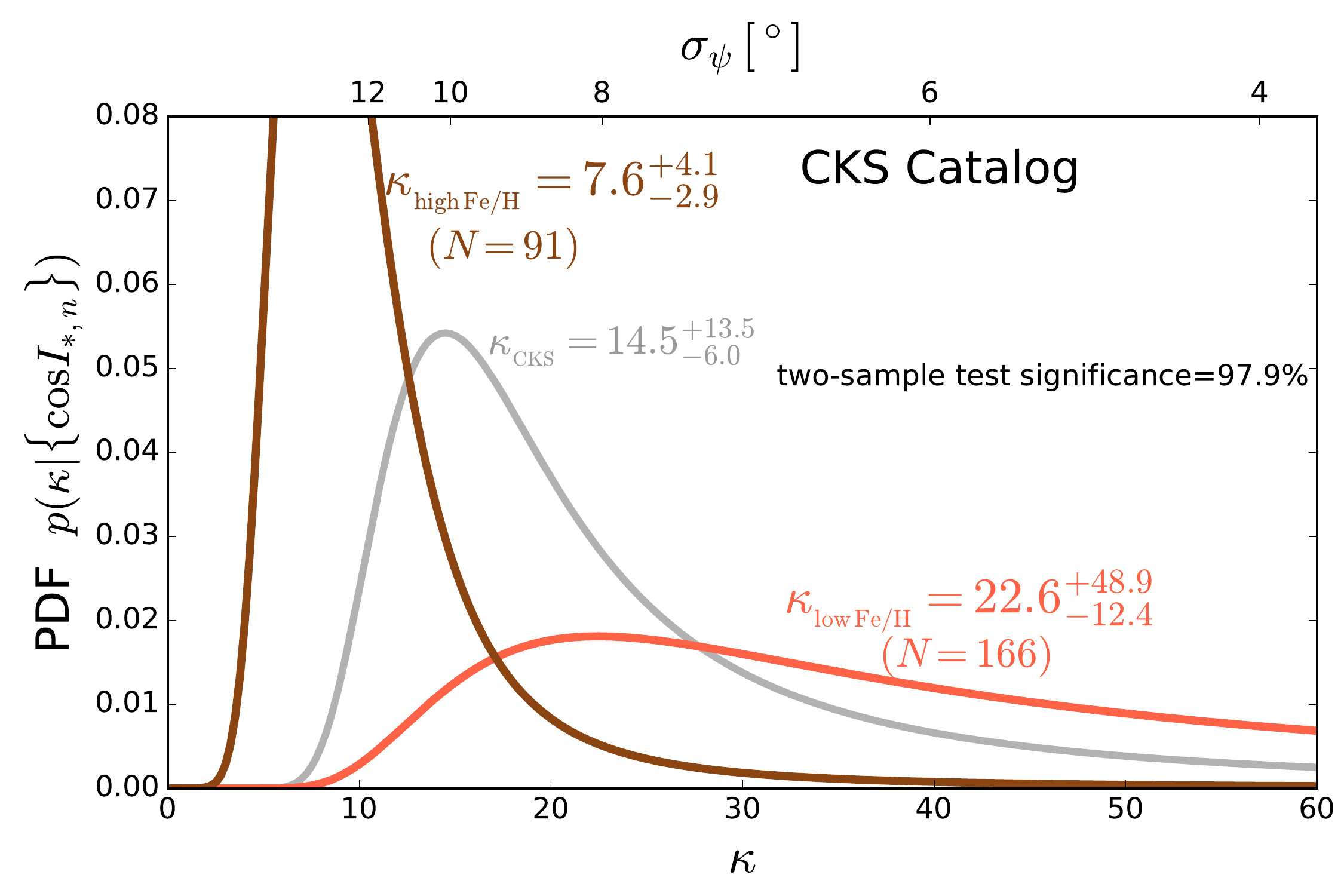}
\includegraphics[width=0.47\textwidth]{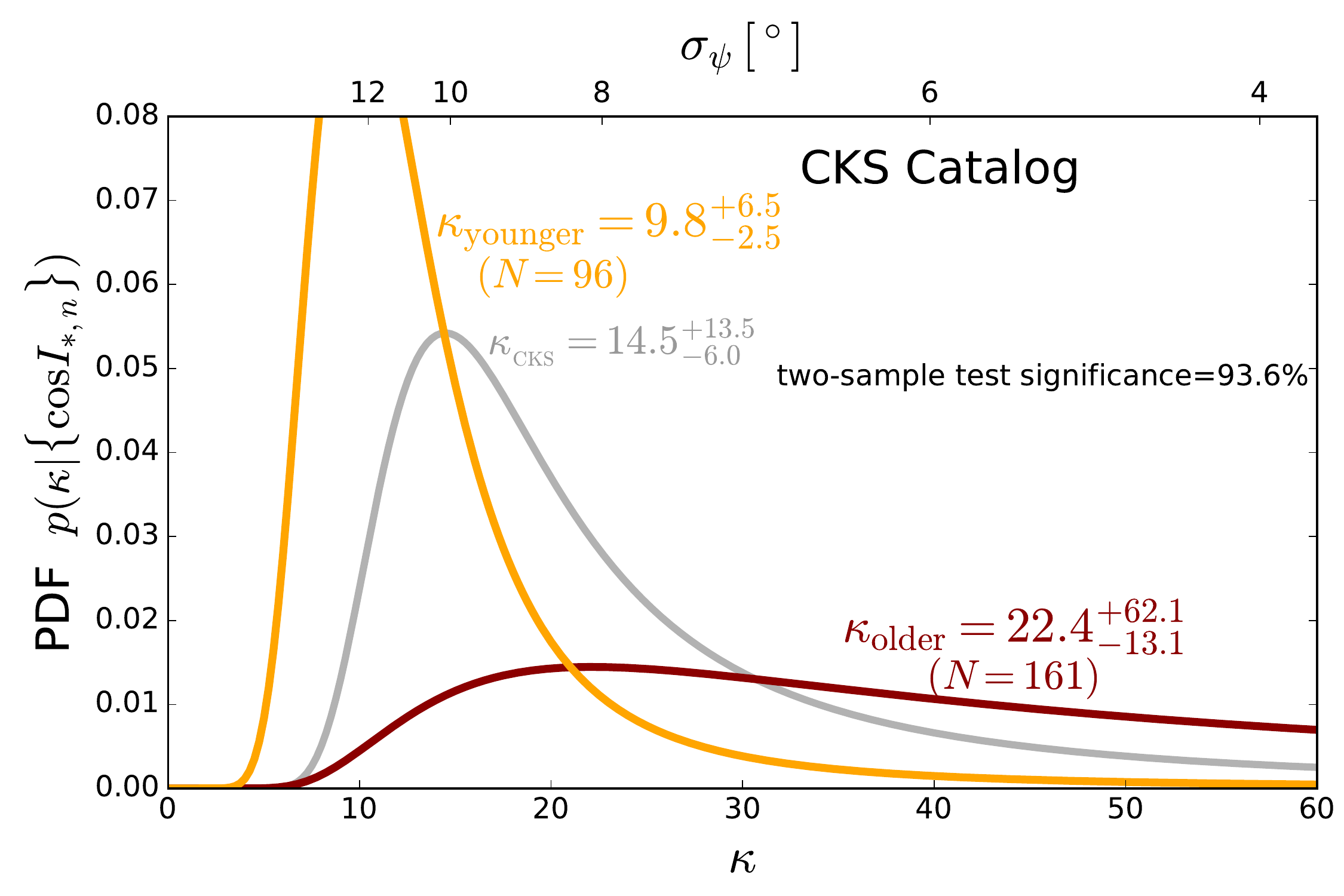}
\vspace{-0.07in}
\caption{Separation of the 257-target dataset into two subsets according to 
stellar metallicity \citep{pet18} and stellar age \citep[ages from][]{joh17} in the CKS survey.
{Top panel: separation onto high-$Z$ systems ([Fe/H]$>=0.135$, $N=$ objects) and
 low-$Z$ systems ([Fe/H]$<0.135$, $N=$ objects). 
 The concentration parameter of the younger subset
is $9.8$ (i.e., $\langle\psi\rangle\approx23^\circ$ and $\sigma_\psi\approx12^\circ$), and
that of the older subset is $22.6$ (i.e., $\langle\psi\rangle\approx15^\circ$ and $\sigma_\psi\approx8^\circ$); the statistical significance of this difference is 97.9$\%$. }
Bottom panel: splitting of the dataset into
younger (yellow) and older (dark red) KOIs \citep[ages from][]{joh17}.  The age cutoff chosen to separate the dataset is $\log A_{\rm cut}=9.63$ (or $A=4.27$~Gyr). The concentration parameter of the younger subset
is $9.8$ (i.e., $\langle\psi\rangle\approx23^\circ$ and $\sigma_\psi\approx12^\circ$), and
that of the older subset is $22.4$ (i.e., $\langle\psi\rangle\approx15^\circ$ and $\sigma_\psi\approx8^\circ$); the statistical significance of this difference is only 93.5$\%$
(a 1.85-$\sigma$ detection). The overlap between the low-metallicity sample ($N=166$) and the older sample ($N=161$) is significant
but not overwhelming; 103 objects satisfy both [Fe/H]$<0.135$ and $\log A>9.63$.
\label{fig:kappa_post_age}}
\end{figure}

\paragraph{Metallicity}{ The high precision provided by the CKS survey allows us to explore the impact of other fundamental stellar properties
such as metallicity and age. In principle, these two variables are not entirely independent from each other, as we expect the older
sample to have somewhat lower metallicity than the younger sample. However, the dynamical range in age is too narrow to reveal
any correlation with metallicity; furthermore, both the lowest and highest metallicities in the sample, [Fe/H]$=-0.4\pm0.04$ for KOI-623
and [Fe/H]$=0.396\pm0.04$ for KOI-941, have ``old'' ages of $\log A=9.95_{-0.13}^{+0.08}$ and $\log A=10.03_{-0.27}^{+0.12}$ respectively.
Fig.~\ref{fig:kappa_post_age} (top panel) shows the concentration inference for a separation of the dataset into a low-metallicity sample ($N=166$)
and a high-metallicity one ($N=91$). The cut in metallicity is [Fe/H]$_{\rm cut}=0.135$, and is chosen as the one that maximizes the difference
in the obliquity properties of the two subsamples, {i.e., the value of cutoff is a free parameter}, which is explored in a ``sweeping'' fashion. Inference results in a concentration parameter of
$\kappa^{\text{\tiny CKS}}_\text{\tiny low [Fe/H]}=22.6^{+48.9}_{-12.4}$ for the lower metallicity subsample, and 
$\kappa^{\text{\tiny CKS}}_\text{\tiny high [Fe/H]}=7.6^{+4.1}_{-2.9}$ for the higher metallicity one. 
The two PDFs appear different,
and thus a proper statistical assessment of their true distinctiveness is required. Following MW14,
we use the squared Hellinger distance $\delta_{{\rm H}^2}$ as a difference metric between
the two distributions. Then, we repeat the hierarchical inference 5000 times by Monte Carlo sampling the dataset in a way that two samples $D_1$ and $D_2$ 
of sizes $N_1=166$ and $N_2=91$ are generated at random
for each try. For each of these tries, we compute $p(\kappa_1|D_1)$ and $p(\kappa_2|D_2)$, and measure the corresponding $\delta_{{\rm H}^2}$. We then count the fraction of realizations in which 
the synthetic $\delta_{{\rm H}^2}$ is equal or larger than in the real sample. This test quantifies
the likelihood of obtaining the observed difference between  $\kappa^{\text{\tiny CKS}}_\text{\tiny low [Fe/H]}$ and 
$\kappa^{\text{\tiny CKS}}_\text{\tiny high [Fe/H]}$ by mere chance. 
We find that in $\sim2\%$ of the random
tries, the two resulting PDFs are more different than in the top panel of Fig.~\ref{fig:kappa_post_age} (by measure of $\delta_{{\rm H}^2}$), thus
concluding that this difference is $98\%$ significant (i.e., a 2.3-$\sigma$ detection)  and thus suggestive, but not conclusive. 
Difference in the planet-bearing frequency as a function of metallicity have been reported in the literature \citep[e.g.,][]{mul16,pet18},
and thus the (moderate) trend of $\kappa$ with [Fe/H] might be an indirect reflection of a dependence of $\kappa$ on planet type (see 
Section~\ref{sec:planet_properties} below).
For example, one might expect, qualitatively, that larger and more numerous rocky cores are formed in high metallicity protoplanetary
disks \citep{pet18}. The lower values of $\kappa$ at higher metallicities might be linked to the formation of more crowded/tightly packed systems,
that will tend to be less stable \citep[e.g.][]{pu15}, thus evolving toward excited mutual inclinations and obliquities.}

\paragraph{Age} Stellar obliquity can be affected 
by stellar spin-down as well as by tidal coupling to close-in planets \citep[e.g.][]{win10,daw14,alb13,li16}, both of which
act over long periods of time. Thus, provided that accurate estimates and a wide enough ranges
of stellar ages are available, one can in principle probe for changes in $\kappa$ as a function of 
this quantity  \citep[e.g., see][]{tri11}.
 As a part of the CKS, \citet{joh17} fitted evolutionary models to the observed spectroscopic parameters to obtain stellar masses and ages. We add theses ages to the 257
target samples and split the dataset into an ``older'' subset ($\log A\geq\log A_{\rm cut}=9.63$)
 with $N=161$ targets and a ``younger'' subset ($\log A<\log A_{\rm cut}$)  with $N=96$ targets, where the cutoff
value was deliberately chosen as the one that maximizes the statistical significance of the data
splitting. We show the results of the concentration inference as a function of age in 
Fig.~\ref{fig:kappa_post_age} (bottom panel). We find that the youngest systems are
consistent with a distribution of obliquities with 
$\kappa^{\text{\tiny CKS}}_{\rm younger}=9.8^{+6.5}_{-2.4}$. 
The ``older'' subset produces 
$\kappa^{\text{\tiny CKS}}_\text{older}=22.4^{+62.1}_{-13.1}$.
This difference has a statistical significance that is marginal at best
($93.6\%$), and further studies will help decide whether this trend is real or not. {As with the metallicity cutoff, 
the age cutoff is a free parameters, which we explore systematically, always requiring that both subsamples had more than 20 objects.}

\begin{figure}[t!]
\includegraphics[width=0.47\textwidth]{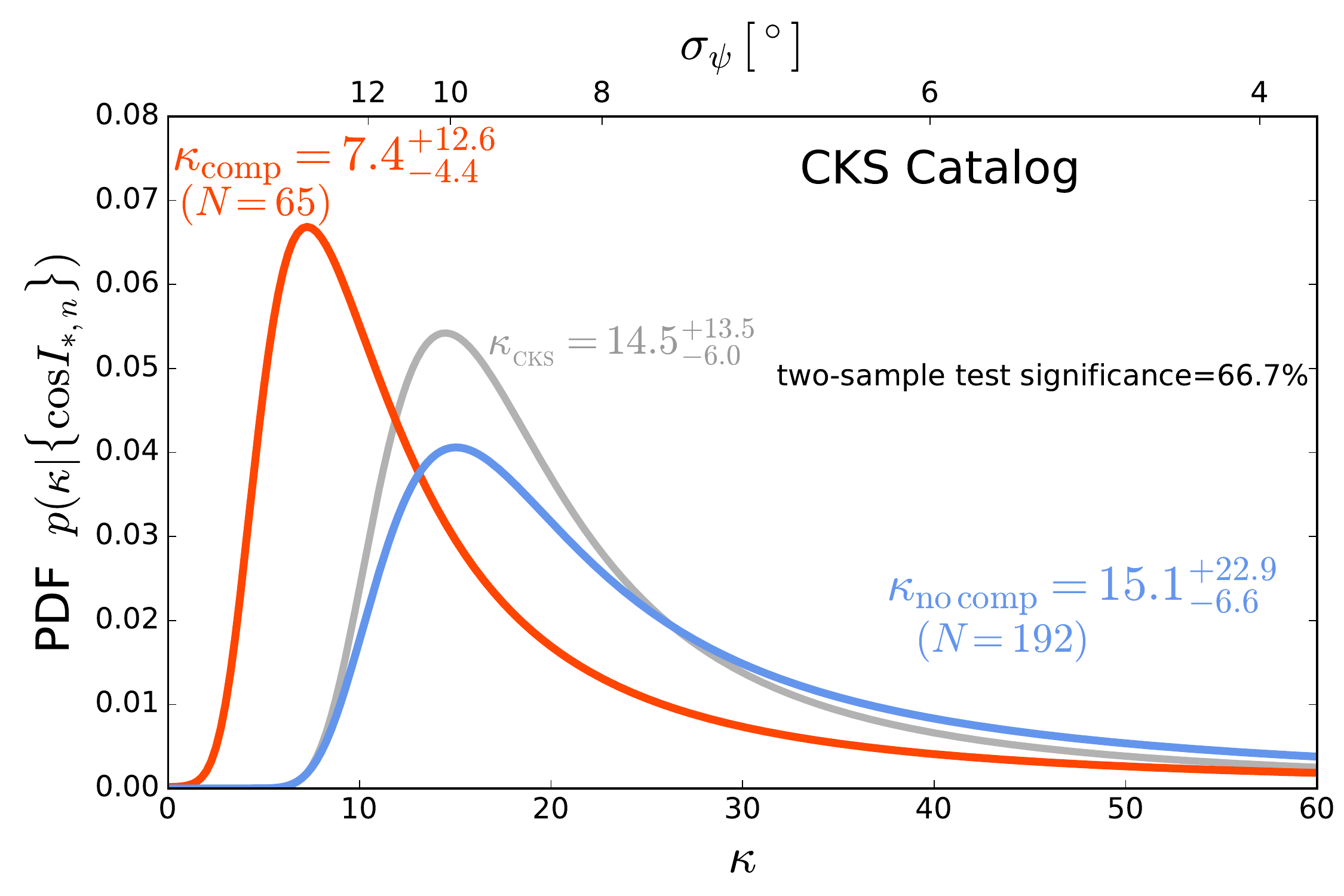}
\vspace{-0.07in}
\caption{Separation of the 257-target dataset into two subsets according to the
presence ($N=65$) or absence ($N=190$) of stellar companions. Planet hosts with  companions
(red curve)
appear to correspond to a lower value of $\kappa$ (a wider distribution of obliquities)
than companionless systems (blue curve), but the
statistical significance of this difference is negligible (it is $46\%$ likely to have happened by
a random separation of the dataset). 
\label{fig:kappa_post_companions}}
\vspace{-0.1in}
\end{figure}

\paragraph{Stellar Multiplicity} Another interesting property that can affect the obliquity of {\it Kepler} systems is  
stellar multiplicity. Different proposed channels for the origins of hot Jupiters
\citep[see][for a recent review]{daw18} 
typically invoke the presence of a distant companion of stellar or planetary mass that
triggers the Lidov-Kozai mechanism or some related secular process responsible for
high-eccentricities \citep[e.g.][]{nao16}.
Some of these studies have explicitly 
provided predictions on the distribution
of obliquities generated \citep[see, e.g.,][]{fab07,nao12,pet15a,pet15b,and16}. However,
moderate obliquities can be induced by a companion whether a system harbors a hot Jupiter or
not, and these companions are not required to be in highly inclined orbits \citep[e.g.][]{bai16,lai16,lai18}.
To test this, we use the adaptive optics follow-up observations of \citet{fur17}. This campaign -- part of the {\it Kepler}
Follow-up Observation Program KFOP (see Section~\ref{sec:other_catalogs} below -- observed
3357 KOI host stars, reporting on the detection of 2297 companions to 1903 stars. Of the 773 KIC stars in the CKS catalog 
with confirmed planets (Section~\ref{sec:CKS}), 766 were part of the observation log of \citet{fur17}, including all of the 257 targets
used in the present work.
 Using the results of the high-resolution imaging campaign of \citet{fur17}, we
assign companion numbers to our 257 CKS targets. We find that 65 KOIs have one or more
stellar companion candidates. Of these, 9 have  $I_*\neq90^\circ$ at a statistically significant level.
As before, we split the dataset according to the presence of
companions and carry out the concentration inference for each subsample. \
In Fig.~\ref{fig:kappa_post_companions} (top panel) we show the posterior PDFs corresponding to
each subset, finding that
 $\kappa^{\text{\tiny CKS}}_{\rm comp}=7.4^{+12.6}_{-4.4}$ and
$\kappa^{\text{\tiny CKS}}_\text{no comp}=15.1^{+22.9}_{-6.6}$. 
Although the $\kappa$ posteriors appear somewhat distinguishable,
the statistical significance of this difference is negligible, as the synthetic distance metric
$\delta_{\mathrm H^2}$ was larger than the measured one in $\sim 35\%$ of the random data splittings attempted.
The undetected effect of stellar multiplicity on obliquity might not be a surprise. The majority of these companions are detected
beyond separation of $1.5''$, which in most cases (assuming a mean distance of $\sim500$~pc) corresponds to 750 AU, a separation
that might be too wide for a significant obliquity to accumulate due to differential precession over the age of the system 
\citep[see, e.g.,][ and the \nameref{sec:discussion} below]{bou14b}. Note that we have removed 34 targets with close companions for being possible blends that can make the period detection ambiguous (Section~\ref{sec:CKS}). We have repeated the obliquity inference reinstating these KOIs
to see if they enclose some tentative clues regarding the influence of companions. Again, the $\kappa$ posteriors of the ``companions'' and
''no companions'' subsamples are statistically indistinguishable.

Since the size of the sample allows us to split the dataset according to more than one variable, 
we have also explored planet multiplicity in combination with stellar multiplicity. No convincing trends arise,
although some multis seem misaligned at a statistically significant level. This is the case KOI 377 (3 transiting planets) and KOI 1486 (2 transiting planets), for which the 95$\%$-confidence upper limits on $\cos I_*$ are 0.108 and 0.143, respectively (see caption in Fig.~\ref{fig:cosipdf_cks}).
This might seem counterintuitive at first, since the presence
of additional planets could protect the system against external perturbations \citep[e.g.][]{bou14b, lai17,lai18}.
However, multi-planet systems could still shield each other from 
the excitation of {\it mutual} inclinations, while still
being coherently inclined with respect to the stellar spin axis. The precise balance between external excitation and suppression of inclinations
depends on all semi-major axes and masses of a given system, and
thus any distinction between oblique and non-oblique populations might be much more subtle than
what we have attempted here.

\subsubsection{Obliquity Trends: Planetary Properties}
\label{sec:planet_properties} 
As an initial test, in Section~\ref{sec:obliquity_distribution} we explored the role of planet multiplicity. However, we can explore
other planetary properties such as orbital period and planet radius.

\begin{figure}[h!]
\includegraphics[width=0.44\textwidth]{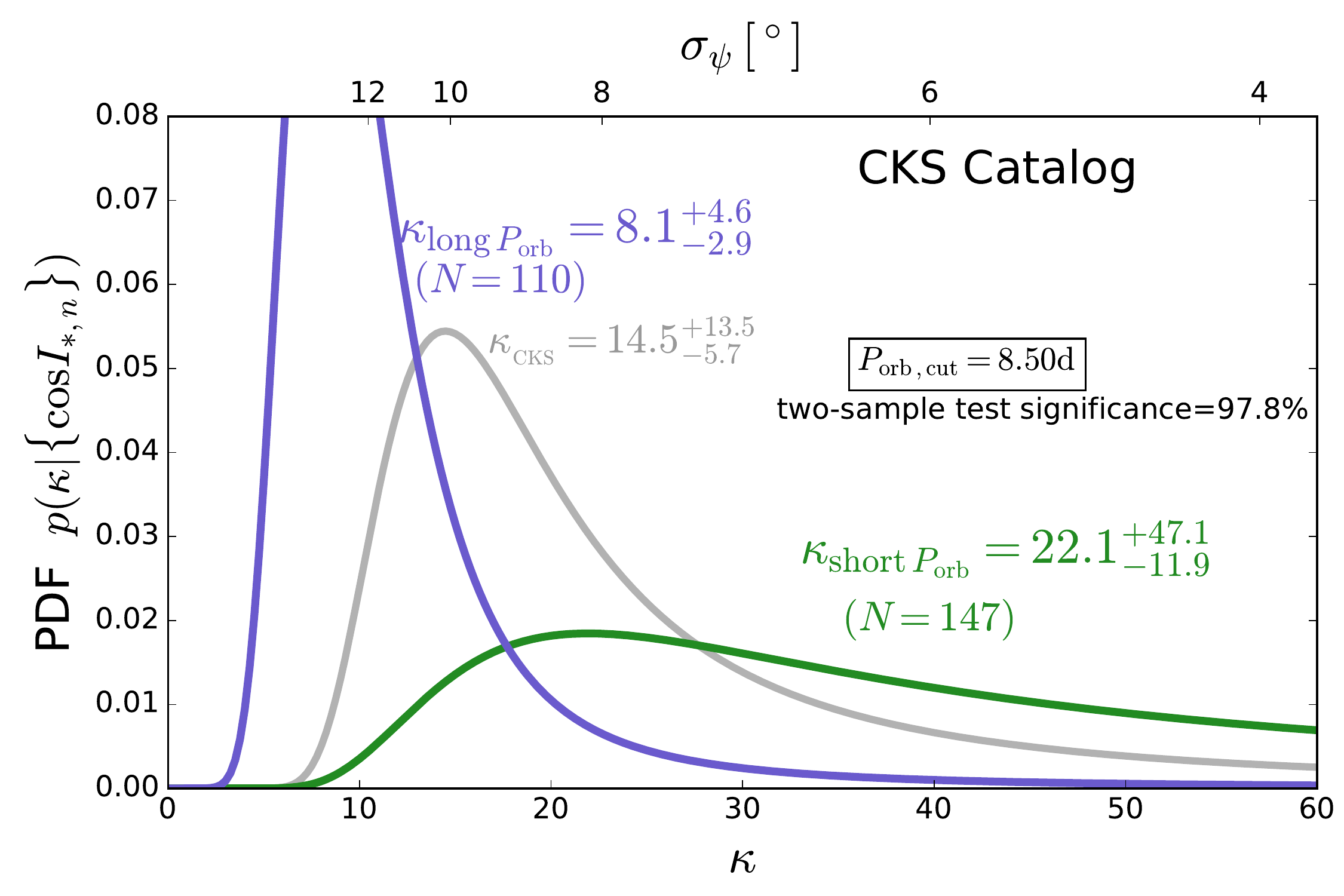}
\includegraphics[width=0.44\textwidth]{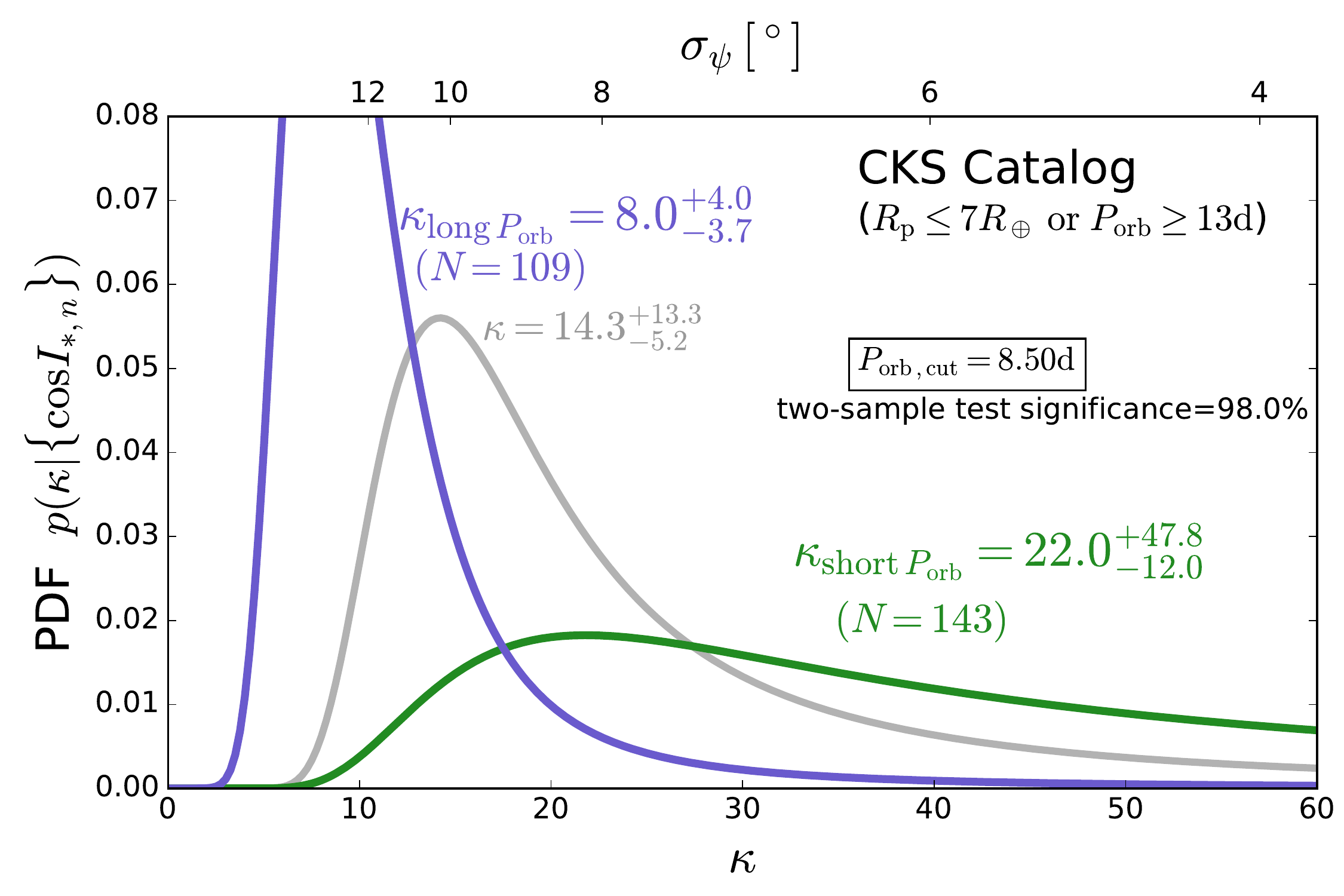}
\vspace{-0.1in}
\caption{{Similar to figs.~\ref{fig:kappa_post_age} and \ref{fig:kappa_post_companions}, this time
separating the dataset by the orbital period of the closest-in planet. Top panel: separation
of the entire 257-target into shorter-period systems ($P_{\rm orb}\leq8.5$~d) and longer-period
systems ($P_{\rm orb}>8.5$~d), producing 
$\kappa^{\text{\tiny CKS}}_{{\rm long}\,P_\text{\tiny orb}}=8.1^{+4.6}_{-2.9}$ (more oblique) and 
$\kappa^{\text{\tiny CKS}}_{{\rm short}\,P_\text{\tiny orb}}=22.1^{+47.1}_{-11.9}$ (less oblique).
Bottom panel: same as above, but after excluding the 5 targets that can be classified
as hot Jupiter systems ($R_{\rm p}> 7R_\oplus$ and $P_{\rm orn}<13$~d). The obliquity trends are unaffected
after removing those targets, and there is a slight increase in the statistical significance of the difference between the two subsets.
In this case, the gray curve corresponds to the total sample after removing the 5 hot Jupiter targets.}
\label{fig:kappa_post_period}}
\end{figure}

\paragraph{Orbital Period} {
Under the tidal-evolution hypothesis of \citet{win10} and \citet{alb12}, stellar obliquity should exhibit some dependence on planetary
orbital period $P_{\rm orb}$. Using the photometric amplitude as an indicator of mean obliquity,  \citet{maz15} concluded that 
low obliquities around cooler KOIs can extended out to $\sim50$~days. However, using the same dataset,
 \citet{li16} argue that the evidence for spin-orbit alignment partially weakens for $P_{\rm orb}\gtrsim10$ days, and that it disappears
when $P_{\rm orb}\gtrsim30$~days. In Fig.~\ref{fig:kappa_post_period} (top panel), we show the splitting of the CKS dataset 
by the period of the closest-in planet reveals a difference between short-period systems and long-period ones. For a period cutoff of
$P_{\rm orb,cut}=8.5$~days, we find that short-period systems are more spin-orbit aligned that long-period ones, and that the statistical
significance of this difference is of $97.8\%$ (a $\lesssim$2.5-$\sigma$ detection). We remove the 5 targets that can be classified
as hot Jupiters and repeat the statistical test (bottom panel of Fig.~\ref{fig:kappa_post_period}), obtaining a mild increase in
the significance of the trend. Despite the small number of targets removed, this additional test is not superfluous, as these 5 targets are
the best candidates to test the tidal realignment hypothesis of \citet{win10} \citep[see also][]{alb12,daw14}. The fact that closer-in planets
are, on average, more spin-orbit aligned with their host stars is in qualitative agreement with the tidal realignment-hypothesis; however,  as
noted by \citet{li16}, it is suspect that this trend  still applies for the small-mass planets in the CKS survey (see \nameref{sec:discussion}).}

\begin{figure}[t!]
\includegraphics[width=0.46\textwidth]{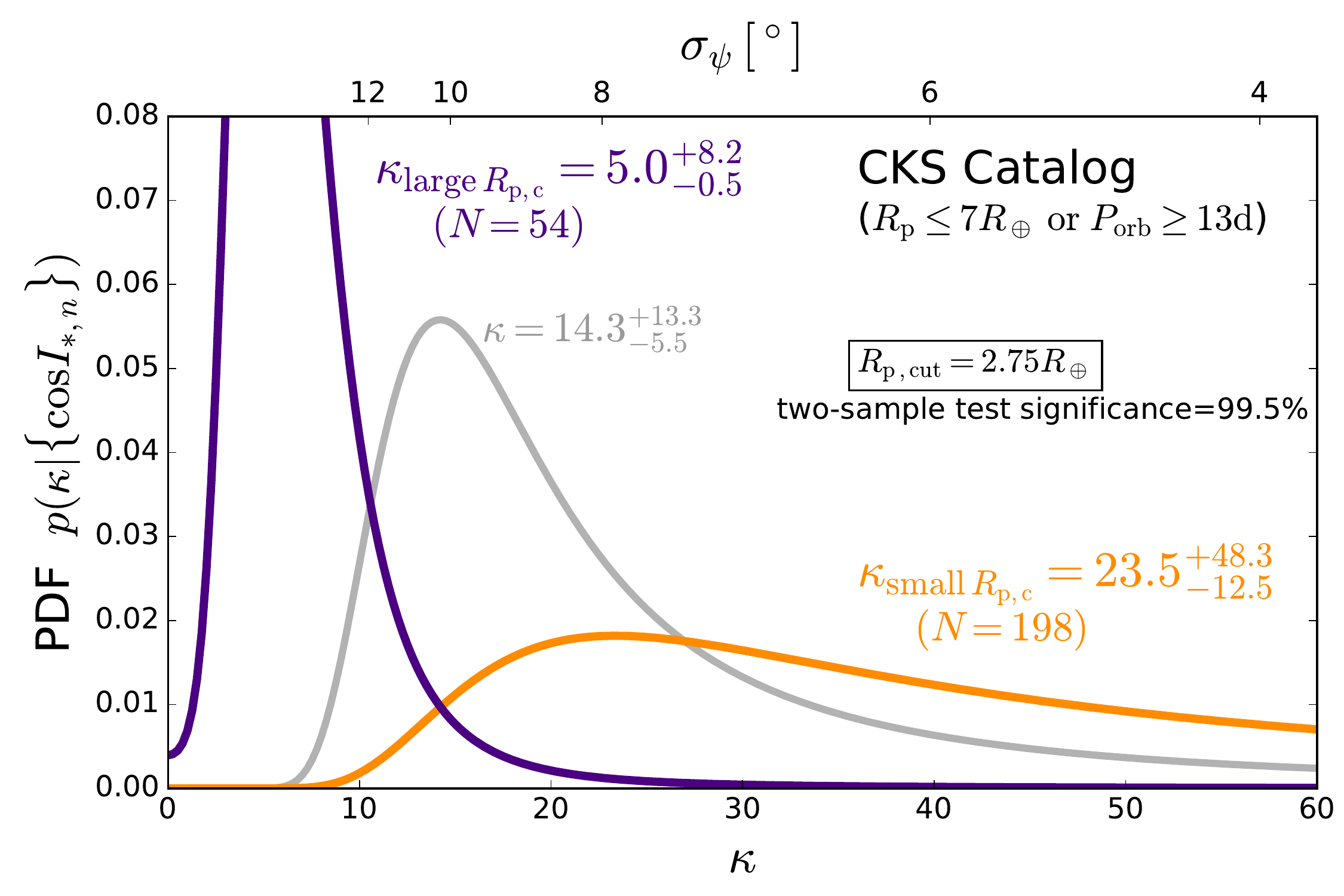}
\includegraphics[width=0.46\textwidth]{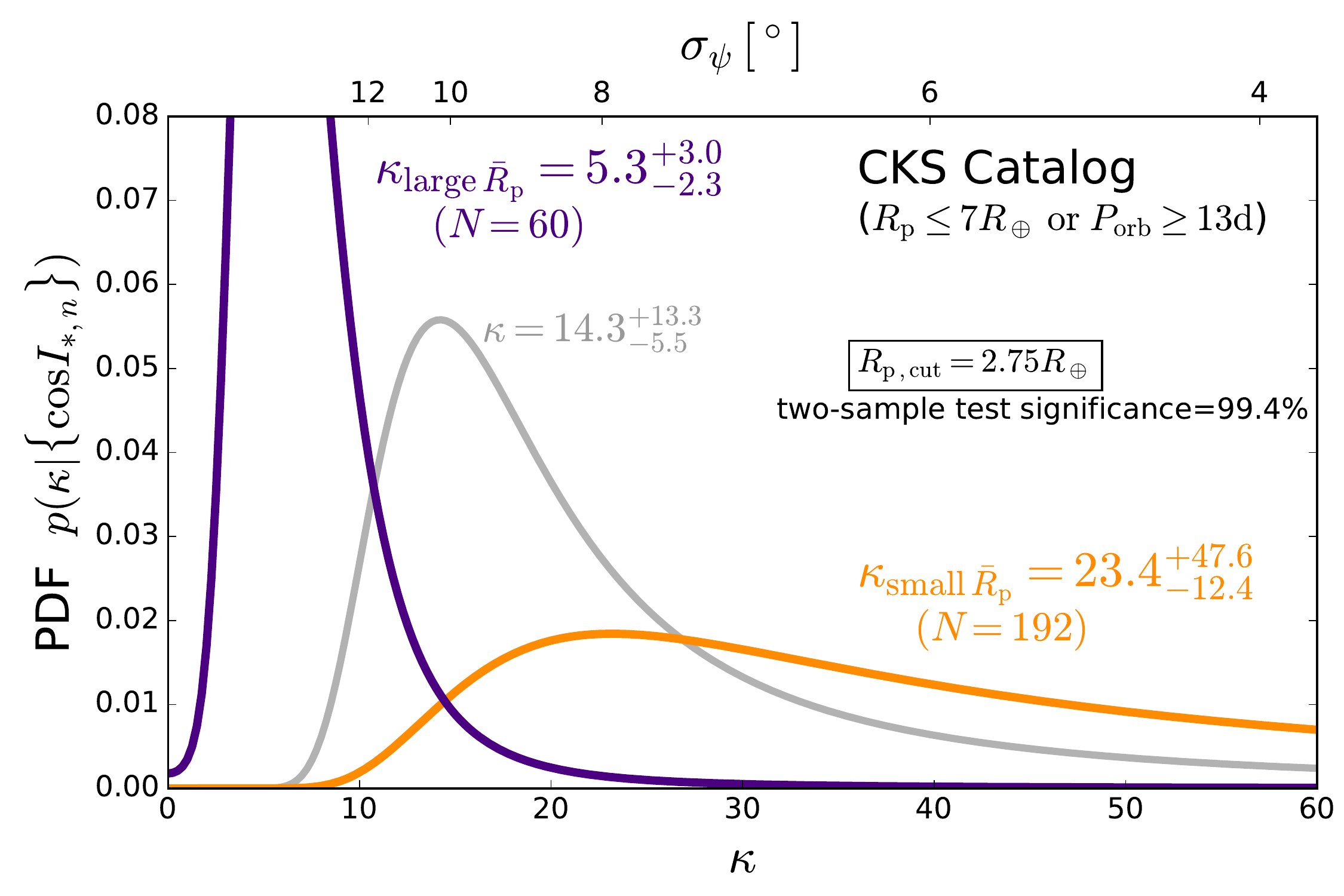}
\includegraphics[width=0.46\textwidth]{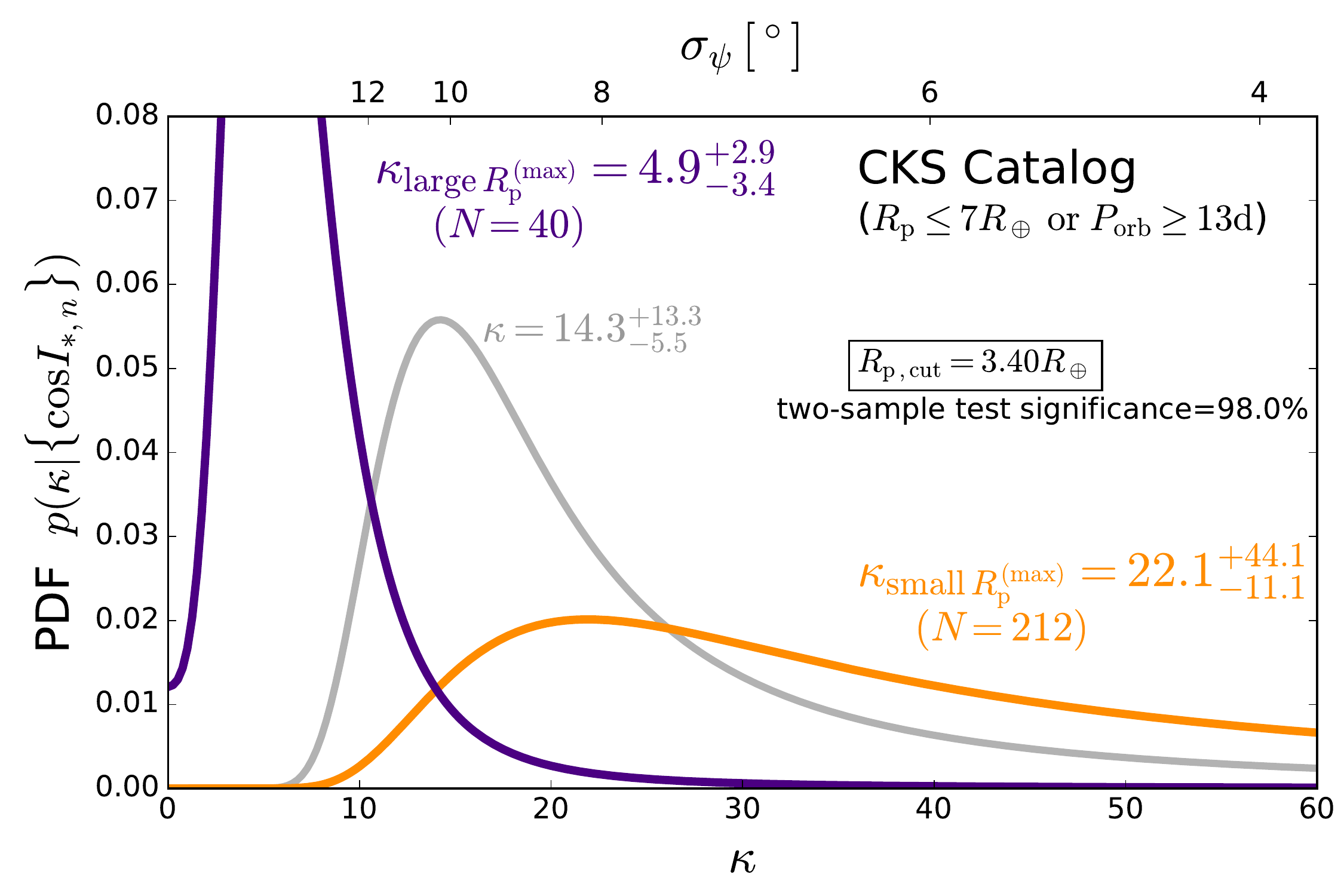}
\caption{{Similar to Figs.~\ref{fig:kappa_post_age}-\ref{fig:kappa_post_period}, this time separating
the CKS dataset (minus five hot Jupiter systems) by three different metrics related to the size of the planets in each system.
Top panel: separation of the dataset by the radius of the closest-in planet $R_{\rm p,c}$, where the radius cut is
placed at  $R_{\rm p,cut}=2.75\,R_\oplus$; the large planet sample is more oblique on average with a concentration parameter
$\kappa_{{\rm large} R_{\rm p}}=5^{+8.2}_{-0.5}$ and the small planet sample is less oblique
with $\kappa_{{\rm small} R_{\rm p}}=23.5^{+48.3}_{-12.5}$.
Middle panel: separation of the dataset by $\bar{R}_{\rm p}$, the mean planet radius, with 
 $R_{\rm p,cut}=2.75\,R_\oplus$; as  before, the large planet sample is more oblique than the small planet sample:
 $\kappa_{{\rm large} R_{\rm p}}=5.3^{+3}_{-2.3}$ and  $\kappa_{{\rm small} R_{\rm p}}=23.4^{+47.6}_{-12.4}$.
 Bottom panel: separation of the dataset by maximum planet radius in each KOI ${R}^{\rm(max)}_{\rm p}$, with 
 $R_{\rm p,cut}=3.4\,R_\oplus$; the large planet sample is more oblique than the small planet sample:
 $\kappa_{{\rm large} R_{\rm p}}=4.9^{+2.9}_{-3.4}$ and  $\kappa_{{\rm small} R_{\rm p}}=22.1^{+44.1}_{-11.1}$.
 The two-sample statistical significance is $99.5\%$ (almost 3-$\sigma$) for $R_{\rm p,c}$
 and $\bar{R}_{\rm p}$, and $98\%$ for ${R}^{\rm(max)}_{\rm p}$.}
\label{fig:kappa_post_radius}}
\end{figure}

\paragraph{Planet Radius} {
For each KOI, we define three statistics: the radius of the closest-in planet $R_{\rm p,c}$,
the mean planet radius $\bar{R}_{\rm p}$ and the radius of the largest planet
 $R_{\rm p}^{\rm (max)}$. As with the exploration of orbital period, we split the dataset
 along these three quantities, sweeping the cut value $R_{\rm p,cut}$ until we find a maximum
 in the statistical significance. The results of this exploration are presented in Fig.~\ref{fig:kappa_post_radius},
where the five hot Jupiter systems have been excluded in all three tests. 
 All data splitting tests provide a consistent trend:
 systems with larger planets are more oblique than systems with smaller planets. For the metrics
 $R_{\rm p,c}$ and $\bar{R}_{\rm p}$ (top and middle panels of  Fig.~\ref{fig:kappa_post_radius}),
  the statistical significance of this trend is  $99.5\%$ ($\lesssim3{-}\sigma$). 
 The test suggests that systems containing planets larger than $\sim3R_\oplus$ (i.e., Neptunes and sub-Saturns)
have larger stellar obliquities. MW14 discussed the possibility of this trend being behind their reported
dependence on planet multiplicity, as multiple-transiting systems tend to have smaller mean radius  \citep[e.g.,][]{lat11}.
This idea is supported by the fact that the planet multiplicity trend is not present in the CKS catalog (see Fig.~\ref{fig:kappa_posterior_cks}),
 while the dependence on  planet radius
is substantial. In conjunction with the stellar metallicity trend (Fig.~\ref{fig:kappa_post_age}), the obliquity-planet radius trend
 appears to point toward a dependence on {\it the total mass contained in planets}, in turn a measure of the ``dynamical
temperature'' of a system \citep[e.g.][]{tre15}.  As we do not have accurate mass estimates
 for most of the planets in this sample, we cannot confidently define a statistic for the total mass contained in planets; however,
 in the \nameref{sec:discussion}, we speculate on ways of approximately assigning a total planetary mass to each KOI.}
 
 \vspace{0.1in}
 {
We briefly comment on the ``sweeping data slicing'' that we implemented above  (Sections~\ref{sec:stellar_properties}  and~\ref{sec:planet_properties} )in order to identify potential breaks in the obliquity distribution.
 This technique was necessary because, for the variables tested, we lack a hypothesis
predicting a change in the distribution of obliquities at some critical value of the variable in question.
 In addition, these variables -- such as metallicity, age, planet orbital period and planet radius -- are continuous, as opposed
to categorical -- such at multiplicity or planet type -- which makes the a priori identification of an adequate cutoff ambiguous.
}

{
The sweeping of cutoff values means that we are carrying $N$ different hypothesis tests, one for each time a new cutoff is tried
and the dataset is split. For each test, the ``null hypothesis'' $\mathcal{H}_{0,n}$ ($n=1,..,N$) is that the two subsets resulting from the splitting are indistinguishable from each other;
we look for the $n$-th cutoff for which we can reject the null hypothesis at some confidence level.
If the properties of the obliquity dataset change abruptly at the $n$-th cutoff,
then the $n$-th $p$-value will be much smaller than that of any of the other tests (as it would be expected to happen, say, at the Kraft break). Alternatively, the obliquity data could depend smoothly and monotonically
on some of these continuous variables, in which case the $p$-value could be small, not only for one, but for many of the $N$ tests.
}

{
When such multiple tests are attempted, it is important to quantify whether the null hypothesis in a given one of them could be rejected ``by chance'' due to  ``$p$-hacking''. 
This is sometimes referred to as the ``multiple comparisons problem'' \citep[e.g.,][]{mil81}.  Multiple strategies have been
devised to correct for this effect, the most widely used being the Holm-Bonferroni method \citep{hol79}, which adjusts the $p$-values of each test\footnote{
In this case, the $p$-value of each test is $p=1-s$, where $s$ is the significance in the nomenclature we have used.} 
based on the number $N$ of tests carried out. An obstacle in implementing this adjustment is the clear correlation of our $N$ tests and of their resulting 
$p$-values \citep[e.g.,][]{con07}. Given this difficulty, we stick to our Monte Carlo approach. This time, we compute statistical significance 
by comparing the distance metric $\delta_n$ resulting from the $n$-th cutoff to the distance metrics obtained from {\it all} cutoffs attempted; thus, we compare to
$5000\times N$ distance metrics with varying cutoff, instead of 5000 metrics with a fixed cutoff. For
example, in Fig.~\ref{fig:kappa_post_age} (top panel), the distance metric of the data splitting according to a metallicity cutoff of [Fe/H]$=0.135$ is $\delta_{\rm H^2}=0.563$,
which is larger than in $96.85\%$ of {\it all} Monte Carlo data splittings across all $N$ tests: thus, the ``global $p$-value'' is 0.0315 (up from
a value of $1-0.979=0.021$ obtained with the previous test); this allows
us to reject the null hypothesis (that there is no metallicity dependence) at an $\alpha=0.04$ level. We repeat this analysis for orbital period   
(Fig.~\ref{fig:kappa_post_period}, top panel), and find a global  $p$-value of 0.0256 (up from $1-0.978=0.022$). Similarly, for the planet radius dependence (Fig.~\ref{fig:kappa_post_radius}, top panel)
we find a global  $p$-value of 0.0093 (up from $1-0.995=0.005$). Therefore, after introducing the global $p$-value as a way of dealing with the multiple comparisons problem, 
the detection significance remains roughly unaltered.
}
 
\subsubsection{Other Catalogs of  $\,V\sin I_*$ Measurements}
\label{sec:other_catalogs}
Spectroscopic studies in the literature can provide with additional values of $V\sin I_*$ and $R_*$ \citep[e.g., see][]{buc12,hir12,hir14}.
However, the CKS has signified a major leap with respect to previous studies, not only because of the size of its sample, but because of the consistency and uniformity of its data collection and analysis.  In general, we expect CKS to supersede any previously reported spectroscopic analyses.
There are however, significant amounts of KOI data that are
 publicly available via KFOP \citep{fur17}, which has compiled follow-up imaging and spectroscopy of a large number KOIs. From the CFOP database, and as of Feb 20th, 2018, we obtain 858 individual KOIs with reported values and uncertainties of $V\sin I_*$ and $R_*$ by one or more users.
Whenever more than one value of either $V\sin I_*$ or $R_*$ is reported, we perform a weighted
mean of the values and their uncertainties. This database overlaps with the CKS database on 505 targets. Unfortunately, these
reported values are inconsistent -- uploaded by different users, using different rotational broadening fitting methods, applied on spectra obtained  from different telescopes -- and often the secondary by-product of a different type of analysis. Although some consistency is found between CFOP and CKS
(rotational velocities tend to agree for $V\sin I_*<5\,$km$\,$s$^{-1}$, albeit with significant scatter), 
the accuracy and uniformity needed for the statistical analysis in the present work make the CKS catalog the only source can be used with confidence.

\vspace{-0.05in}
\subsection{The TEPCat Catalog}\label{sec:TEPCAT} 
The same kind of hierarchical Bayesian inference can be carried out
for a database of projected obliquity measurements via use of
Eq.~(\ref{eq:lambda_given_kappa}). Using an essentially equivalent method, \citet{fab09} inferred
a value of $\kappa$ from a list of 11 targets with RM observations.
Using their sample, but implementing the formalism summarized in Section~\ref{sec:overview}, we 
obtain $\kappa_{\text{\tiny~FW}}=9.1_{-7.1}^{+69}$, in consistency with the results of that work.

\begin{figure}[t!]
\includegraphics[width=0.48\textwidth]{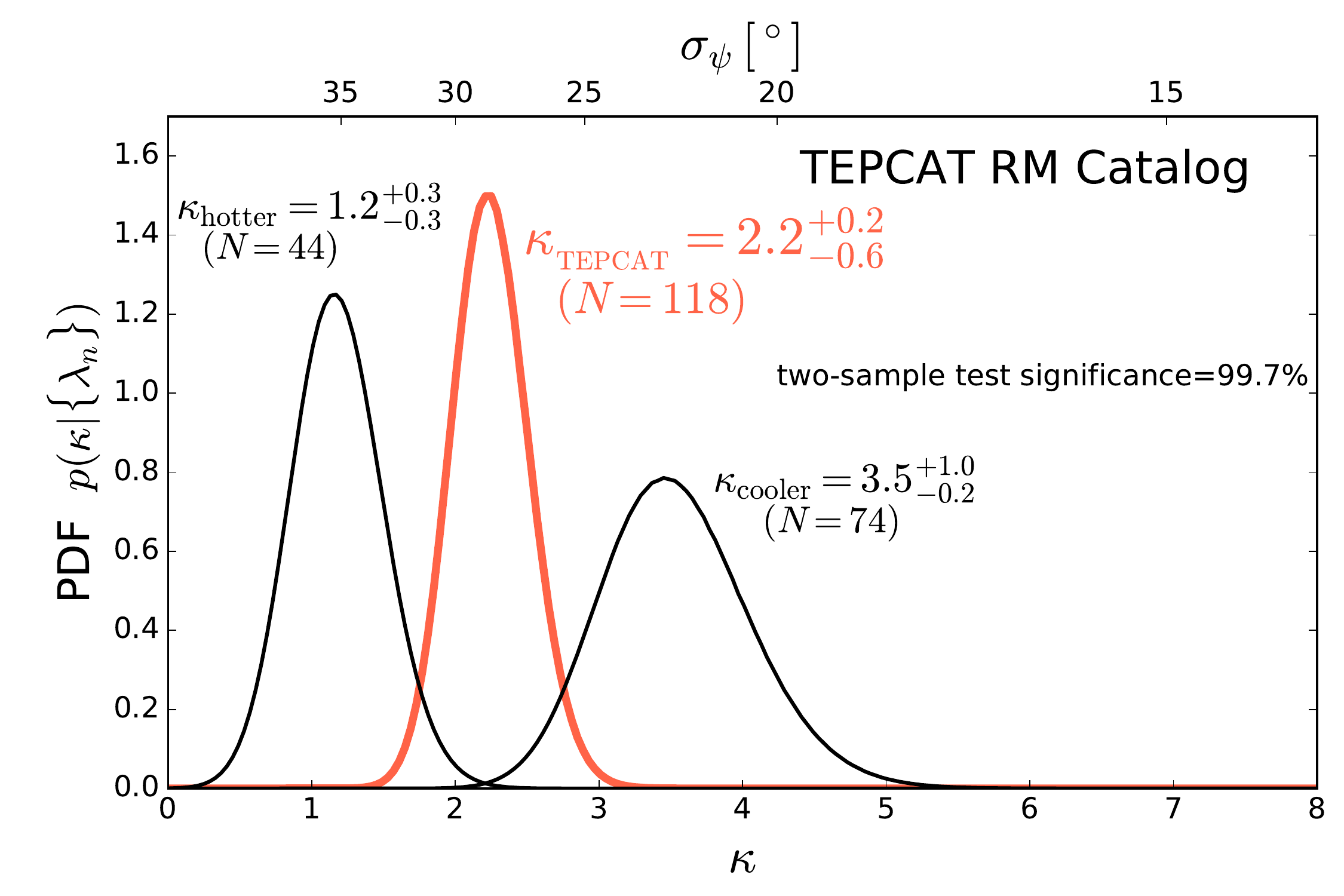}
\includegraphics[width=0.48\textwidth]{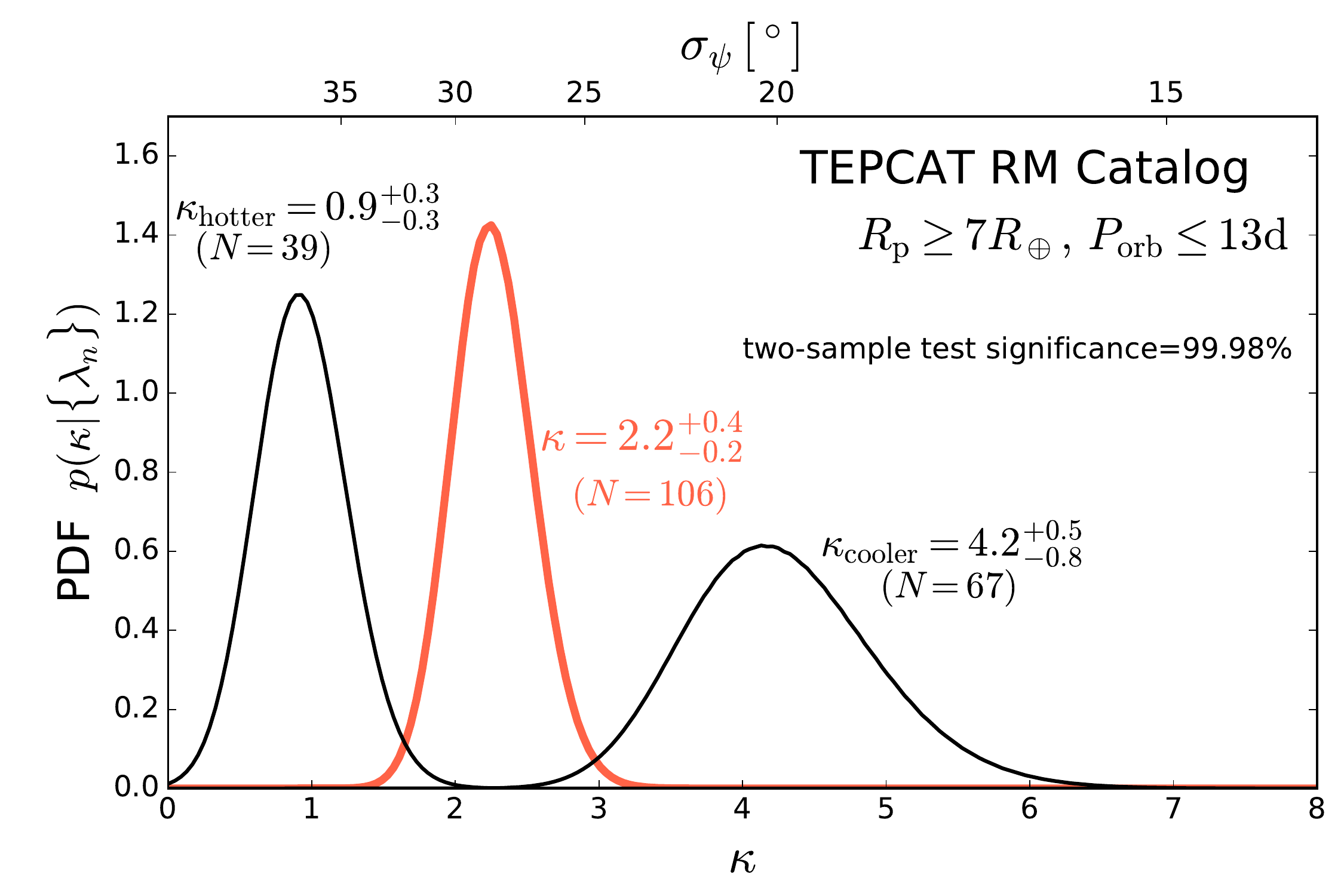}
\vspace{-0.07in}
\caption{Posterior PDF of the $\kappa$ parameter using RM data from the TEPCat catalog \citep{sou11}. The posterior $p(\kappa|{\lambda_n})$ (red curve)  is obtained from the multiplication of Eq.~\ref{eq:kappa_likelihood} by Eq.~\ref{eq:prior_pdf} using Eq.~\ref{eq:kappa_deltalike_lambda}).
Top panel:
The concentration parameter of the entire TEPCat sample (118 objects) is  $\kappa_{\text{\tiny TEPCAT}}=2.2^{+0.2}_{-0.6}$ 
(i.e., $\langle\psi\rangle\approx53^\circ$ and $\sigma_\psi\approx30^\circ$). The dataset
can be divided into hotter and cooler subsamples  ($T_{\rm eff}$ above and below 6250~K)
with the hotter sample being more oblique (lower value of $\kappa$) by a statistically significant
amount (3-$\sigma$ detection). {Bottom panel: same as above, but for those targets in the  TEPCat catalog 
for which  $R_{\rm p}>7R_\oplus$ and $P_{\rm orb}\leq 13$~d (106 objects). In this case,
$\kappa_{\text{\tiny TEPCAT,\,HJ}}=2.2^{+0.4}_{-0.2}$. The splitting of the dataset according to temperature produces $\kappa^{\text{ \tiny TEPCAT,\,HJ}}_{\rm hotter}=0.9\pm0.3$ and $\kappa^{\text{ \tiny TEPCAT,\,HJ}}_{\rm cooler}=4.2^{+0.5}_{-0.8}$,
with the difference being significant at the 4-$\sigma$ level.}
\label{fig:kappa_posterior_tepcat}}
\vspace{-0.1in}
\end{figure}

We can extend the same analysis to a much larger RM database. We retrieve
the data compiled in John Southworth's TEPCat Catalog \citep[][\url{http://www.astro.keele.ac.uk/jkt/tepcat/}]{sou11}. This catalog contains 191 measurements for 118 unique systems. For multiple entries, we take the weighted mean of the observations, provided that
 there is some consistency between
the reported measurements; if discrepancies are found, we take the latest/most accurate measurement. The measurement posterior of $\lambda$ for each target, i.e., $p(\lambda|\tilde{D}_n)$
in Eq.~\ref{eq:kappa_deltalike_lambda}, is taken to be a Gaussian with mean and variance given by each measurement and its uncertainty, respectively. The likelihood of the data given
$\kappa$ is obtained via Eqs.~(\ref{eq:kappa_likelihood}) and~(\ref{eq:kappa_deltalike_lambda}),
and the same prior (Eq.\ref{eq:prior_pdf}) used in Section~\ref{sec:CKS}. The posterior
PDF of $\kappa$ is shown in Fig.~\ref{fig:kappa_posterior_tepcat}, and the $68\%$ probability
interval is given by $\kappa_{\text{\tiny TEPCAT}}=2.2^{+0.2}_{-0.6}$. This value of the concentration
parameter is much smaller than that obtained for the CKS sample, and corresponds to
a much more oblique population ($\kappa=2$ implies $\langle\psi\rangle=53^\circ$ and $\sigma_\psi=30^\circ$). This discrepancy, however, is to be expected, as RM measurements
are typically limited to hot Jupiter systems \citep{gau07}, in contrast to the more diverse nature
of the {\it Kepler} exoplanet systems \citep[see, e.g.,][]{alb12,win17}.

\paragraph{Effective Temperature}
The size of the TEPCat catalog allows us to split the dataset into subsamples. This is
shown in Fig.~\ref{fig:kappa_posterior_tepcat} (top panel), where concentration inference was carried
out for a ``hotter'' ($T_{\rm eff}\geq6250$~K) subset with $N=44$ entries, 
and a ``cooler'' ($T_{\rm eff}<6250$~K) subset with $N=74$ entries. We find that
$\kappa^{\text{ \tiny TEPCAT}}_{\rm hotter}=1.2\pm0.3$ and $\kappa^{\text{ \tiny TEPCAT}}_{\rm cooler}=3.5^{+1}_{-0.2}$ with a statistical significance of $99.7\%$. 
{Although the TEPCat catalog is largely composed on hot Jupiters, not all of its entries
qualify as such. Indeed, if we only select targets with $R_{\rm p}>7R_\oplus$ 
and $P_{\rm orb}\leq 13$~d, we remove 12 targets. Repeating the  separation of the dataset according to $T_{\rm eff}$} (Fig.~\ref{fig:kappa_posterior_tepcat}, bottom panel), 
{we find $\kappa^{\text{ \tiny TEPCAT,\,HJ}}_{\rm hotter}=0.9\pm0.3$ and $\kappa^{\text{ \tiny TEPCAT,\,HJ}}_{\rm cooler}=4.2^{+0.5}_{-0.8}$,
and that the hotter sample and the cooler sample are different with a
$99.98\%$ statistical significance (almost a 4-$\sigma$ detection).}

This obliquity-temperature dependence is in accordance
to the confirmed trend that hotter hot Jupiter hosts tend to be more oblique
cooler ones \citep{sch10,win10,heb11,alb12,daw14,maz15}. Nevertheless, these values of
$\kappa$ indicate that hot Jupiters systems are {\it in general} more oblique
($\kappa^{\text{ \tiny TEPCAT}}_{\rm cooler}\sim 4.2$) than the majority of configurations found in the {\it Kepler} catalog ($\kappa_{\text{\tiny CKS}}\sim14$), regardless of stellar effective temperature. In principle, one could solely focus on hot Jupiters
in the CKS sample to explore the discrepancy between  $\kappa_{\text{\tiny~TEPCAT}}$ and 
$\kappa_{\text{\tiny~CKS}}$. Unfortunately,  only 5/257 targets in the list compiled in 
Section~\ref{sec:CKS} fall into the ``hot Jupiter category'' (see Section~\ref{sec:discussion} below).

\subsubsection{Three-dimensional Obliquities?}
Having PDFs of both $\lambda$ and $\cos I_*$ can,
in principle, be used to construct the three-dimensional obliquity $\psi$ 
\citep[e.g.][]{ben14}. The TEPCat and CKS databases overlap on 5 targets: KOI 377 (Kepler-9), KOI 203 (Kepler-17),
KOI 806 (Kepler-30), KOI 63 (Kepler-63), KOI 94 (Kepler-89). Of these five 
targets, KOI-63 ($T_{\rm eff}=5673$~K) is severely oblique in both $\lambda$ and $I_*$, as already reported by \citet{san13}. KOI 63 is also unusual because it is one of the fastest rotators in the sample
with $P_{\rm rot}=5.5$~days (the mean in the dataset is 18.8~days).

\subsection{Obliquity Distribution of Hot Jupiters in the CKS and TEPCat Datasets}
The significance of the temperature separation of the TEPCat sample is clear (Fig.~\ref{fig:kappa_posterior_tepcat}). However,
we lack the statistics to carry out an analogous test with the hot Jupiters in the CKS target list.
The overall CKS sample used for our analysis ($N=257$) contains 20 stars above
6200~K ($7.8\%$). Of the 257 targets, only 5 are hosts to hot Jupiters, all below 6000~K. By contrast, 
$37\%$ of targets in the TEPCat catalog (mostly hot Jupiter hosts) have $T_{\rm eff}>6200$~K. This severely
limits our ability to find a consensus between what can be inferred from RM measurements
and via the $V\sin I_*$ method. \citet{win17} found a workaround to this limitation, and
using a statistical analysis of $V\sin I_*$ alone \citep[e.g.][]{sch10},  identified 
six systems above the Kraft break that are likely to be oblique: KOI 2, KOI 18, KOI 98, KOI 167, KOI 1117 and
KOI 1852. None of these systems made it to our original 257-target dataset. KOI 98 was removed due
to risk of blending and the other five have inconsistent SA and QPGP periods.
The SA periods of KOIs 2, 18, 1117 and 1852 
fall in a range of 60 to 90 days ({\it very} long for stars above the Kraft break, and likely to be wrong), while the corresponding QPGP periods are between 5 and 25 days, still somewhat long for these
effective temperatures. { If we use the QPGP periods for KOIs 2 (HAT-P-7), 18 and 1117  -- $19.6\pm4$, $24^{+12}_{-4}$ and $10\pm1$ days respectively --
we find that  these KOIs have equatorial velocities $V_{\rm eq}\approx V\sin I_*$, and thus are consistent
with $I_*\approx90^\circ$. In this case, the low $V\sin I_*$  of these objects would result from
them being  anomalously slow rotators for their effective temperatures, and not from being highly oblique. On the other hand, the QPGP period
of KOI 1852 ($6.9\pm0.2$ days) does imply that this system could be severely oblique with $I_*(95\%)=50^\circ$, but it does not harbor a hot Jupiter.
These estimates contradict some of the findings of} \citet{win17} {in regard to the obliquity of hot Jupiter hosts above the Kraft break. 
It is interesting to note that KOI-2 (HAT-P-7) has already been reported to have either a polar or retrograde planetary orbit} \citep{win09b}. 
{Thus, if indeed $\sin I_*\approx1$ for this system, that could only be consistent with a fully retrograde configuration.}

Whenever the SA and QPGP periods differ, it is typically when SA periods are extremely long.
Some SA periods are longer than 100 days, while none of the QPGP periods are longer than 55 days.
These extremely long periods could be an artifact of time series analysis or inherent to the {\it Kepler}
instrumental systematics, which were never optimized to capture such long variability timescales 
\citep[see][]{mon17}. Thus, in the absence of matching, if one rotational period is to be chosen, we favor those of  \citet{ang18}. {A further advantage of these estimates is that
the confidence intervals in $P_{\rm rot}$ of }\citet{ang18} {are
the well-defined result of the Bayesian fitting of a parametric (albeit non-physical) model.} If we use the QPGP for hot Jupiters, we increase the number of such objects
in our database from 5 to 19. Under these circumstances, only four hot Jupiters are significantly
oblique -- KOIs 97, 127, 201 and 214 -- of which only KOI 97 has $T_{\rm eff}>6000$~K. 
The upside of this exercise is that this list of 19 hot Jupiters 
is now marginally large enough for $\kappa$-inference.
Our analysis results in $\kappa_\text{\tiny HJ}=0.01^{+11}_{-0.01}$, with a $\kappa$ posterior that is marginally
distinguishable from the prior $\pi_\kappa$ (Eq.~\ref{eq:prior_pdf}).
If we instead use the prior $\pi'_\kappa$ in 
Eq.~(\ref{eq:prior_pdf_alt}), which we may argue is a better representation of a Jeffreys prior at large 
$\kappa$, we find $\kappa_\text{\tiny HJ}=4^{+33}_{-3.9}$. It is difficult to conclude something from these results alone, but these concentration parameters cannot rule out that $\kappa_{\text{\tiny TEPCAT}}\approx2.2$
provides an adequate representation of the underlying obliquity distribution of hot Jupiters.

\section{Discussion}\label{sec:discussion}

We have explored the influence of seven different variables on the obliquity concentration parameter using
two publicly available catalogs of exoplanet systems: the CKS survey and the TEPCat catalog. The variables tested fall into
the ``stellar properties'' category: (1) stellar effective temperature, (2) stellar age, (3) stellar metallicity and (4) 
stellar multiplicity; or into the ``planetary properties'' category: (5) planet multiplicity, (6) planet orbital period
and (7) planet radius. {From the CKS survey, we have found that metallicity, planet orbital period and planet radius are the variables to which
obliquity is the most sensitive, while effective temperature is not testable using this catalog. Planet multiplicity, on the other hand, is found to have no
significant correlation with stellar obliquity}. The TEPCat catalog is used to find a very strong correlation with effective temperature, in agreement
with similar previous claims in the literature.

In particular, exploration of three of those seven variables, namely $T_{\rm eff}$,  stellar multiplicity and planet multiplicity,
was motivated by previous observational and theoretical studies. All these three tests returned null results in the CKS catalog: those variables
do not correlate with stellar obliquity.

The obliquity dependence on stellar effective temperature
for hot Jupiters is the most robust of the  trends found in the literature.
Unfortunately, the $V\sin I_*$ method is not very effective at probing this relation, since the sample
of KOIs with both $V\sin I_*$ and $P_{\rm rot}$ measurements  is sparse
above 6000~K. Hotter stars tend to be more variable and are also larger/brighter, and thus
typically only large planets around the quietest of these stars can be detected \citep{maz15,win17}.
In turn, stars with little variability cannot be used to infer rotational periods via photometric modulation \citep{mcq14,ang18}.
We have attempted several approaches to isolate the hot Jupiter sample and study its obliquity properties, 
but more data are needed to robustly compute a value of $\kappa$ that can be compared
to the one derived from TEPCat data. An alternative use of {\it Kepler} data to infer $I_*$
is asteroseismology, as done by \citet{cam16}. This method represents
a powerful alternative and complement to the $V\sin I_*$ method when stellar activity is too low to
enable  $P_{\rm rot}$ measurements, in turn permitting
a reliable computation of the power spectrum of non-radial oscillations \citep[e.g.][]{cha11}.
Unfortunately, the 25-target
list of \citet{cam16}  contains only one hot Jupiter --KOI 2, which is oblique and above the Kraft break--
not allowing us to address the temperature-obliquity 
relation reported by \citet{win10}. The fact that, by contrast, we find such a strong detection of temperature
dependence in the TEPCat dataset (a 4-$\sigma$ detection, see Fig.~\ref{fig:kappa_posterior_tepcat})
highlights the fact that the CKS and TEPCat catalog are, in general, probing different planet populations.
 
A dependence of $\kappa$ on stellar multiplicity 
could be detectable in similar datasets, provided that we can identify companions in an adequate manner and that these
are close enough to induce obliquity through cumulative differential precession \citep[e.g.,][]{bou14b}.
As we cannot know if the visual companions of \citet{fur17} are truly bound, and we have removed some of the closest to avoid confusion due to blending, the true effect of stellar multiplicity
in the {\it Kepler} sample remains unknown. Perhaps future missions of nearby planet-hosts such as TESS, complemented with astrometric
information from {\it Gaia} \citep[e.g., see][]{qui16}, will not only improve our $\cos I_*$ estimates, but also help identify truly bound multiple stellar systems. In our analysis,
we find that some multi-planet systems with companions show significant spin-orbit misalignment (in particular KOIs 377 and 1486, which drive the measured low value of $\kappa$ for this sub-population).
It is possible that multi-planet systems are more susceptible to external torquing as they have a large collective quadrupole moment; provided that the multi-planet system
is able to react nearly rigidly to external perturbations, the entire coplanar system can
precess around the global angular momentum vector  at a much faster rate than the host star would. 
\citep[e.g.,][]{kai11,bou14b}.
This process introduces an obliquity, which will depend on the mass, separation and inclination of the stellar companion \citep[e.g.,][]{lai16}.

The tentative dependence on planet multiplicity (MW14) is attractive as it fits into a
picture of exoplanet statistics in which systems can be categorized
according to their ``dynamical temperature'' \citep[e.g.][]{tre15}: multi-planet systems
are ``dynamically cold'', having low mutual inclinations and near-zero eccentricities; 
``dynamically hot'' systems, on the other hand, contains fewer planets and have larger eccentricities 
and mutual inclinations \citep{xie16,zhu18}. However, it is known that single-transiting systems are
not necessarily true single-planet systems \citep{tre12}; indeed,  \citet{zhu18} found that an important fraction of these observed singles exhibit signs of transit-timing variations (TTVs). Thus, even as a scrambled distribution of stellar obliquities would be consistent with dynamically hot systems, identifying which systems in our KOI sample are truly hot is difficult without measured eccentricities and/or TTVs.  Unfortunately, we have not 
found any evidence to support this hypothesis by looking at planetary multiplicity alone.

{Three variables appear to exhibit substantial to significant trends in the CKS catalog: stellar metallicity ($97.9\%$ significance), planet orbital period
($98\%$ significance) and planet radius ($99.5\%$ significance).
The dependence on both stellar metallicity (Fig.~\ref{fig:kappa_post_age}) and planet radius (Fig.~\ref{fig:kappa_post_radius}) 
could be pointing toward an underlying dependence on total mass contained in planets $M_{\rm p,tot}$, or the total
orbital angular momentum contained in planets  $\mathbf{J}_{\rm p,tot}$,
probing the dynamical temperature of planetary systems in a more meaningful way than (observed) planet multiplicity alone. 
The total mass contained in planets $M_{\rm p,tot}$ is difficult do estimate from the current dataset. Even using the empirical mass-radius relation of \citet{wei14}, we still need to leave out 32 system for which that relation is not valid (systems with one or more planets with $R_{\rm p}<4R_\oplus$).
If we simply label those 32 systems as ``heavy'' planetary systems and the rest as ``light'' ones, we indeed get a difference in the concentration
parameter that is significant with a $98\%$ confidence. In the future, better planet mass estimates could shed light on the role of
$M_{\rm p,tot}$ and  $\mathbf{J}_{\rm p,tot}$ on the obliquity of the stellar host.}

{The dependence on planet orbital period is intriguing, especially because it seems to hold for planets of any mass. The work of \citet{maz15} suggest that some level spin-orbit alignment can even extend out to orbital periods of $\sim50$ (our dataset contains only 15 KOIs for which the shortest orbital period is greater than 50 days).  It is difficult to picture a
scenario in which low-mass planets are able to modify the obliquity of their host star out to orbital periods of 10 days, let alone 50. Such planets cannot effectively torque the star as the total angular momentum contained in the planetary orbit is too small compared to that contained in stellar spin \citep[e.g.][]{daw14}, and thus this finding presents a severe challenge to
proposed mechanisms of tidal realignment \citep[e.g.][]{li16}.  An alternative idea, proposed by \citet{mat15} is that spin-orbit re-alignment can be achieved by planet ingestions, 
and that low obliquities are the result of the incorporation of orbital angular momentum onto the star, bringing its spin closer to that of the (nearly coplanar) planetary system. Under this hypothesis, stars above the Kraft break are less susceptible to realignment because they are faster rotators. In principle, this hypothesis should produce a correlation between obliquity and stellar rotational period (for $T_{\rm eff}$ above and below 6000~K). Since we use rotational periods to infer obliquities, we refrain from searching from such a correlation at this moment, as it would produce biased results. However, we encourage future observational work aiming at independently measuring obliquities and rotational periods of planet hosts.}

{Yet another novel hypothesis to explain the relation between obliquity and orbital period is that these planets are simply born within a gas disk that is closely aligned with the stellar spin. This would require the innermost regions of protoplanetary disks  ($r\lesssim 0.1$~AU) to be immune to any external mechanisms that act to induce misalignment
 respect to the stellar equator. In principle, 
this can be achieved by a protostellar analog to the Bardeen-Petterson (BP) effect \citep{bar75}, in which viscous accretion disks can reach
a warped steady-state geometry that transitions into of perfect spin-orbit alignment within some transition radius. The BP 
effect is the result of a competition between the nodal precession of test particle orbits at some frequency $\Omega_p$ and the rate at which
material and angular momentum are resupplied via advection \citep{kum85}. To a very rough approximation, once can identify
the BP transition radius $r_{\rm BP}$ with the distance at which precession and advection balance each other, i.e., when $\Omega_p\sim 1/(\alpha^2 t_{\rm visc})$ \citep{pap83,kum85}, where $\alpha$ is the viscosity parameter and $t_{\rm visc}\sim \alpha^{-1}h^{-2}(r) P_{\rm orb}$ is the viscous time and $h$ is the disk aspect
ratio.
The precession rate due to a rapidly spinning star is proportional to the oblateness quadrupole moment $J_2M_*R_*^2$ and scales with
orbital radius as $\propto r^{-7/2}$.
Alternatively, a magnetized protostar for which the magnetic moment $\boldsymbol{\mu}$ and spin vector are misaligned can also induce precessional torques, which act at a rate proportional
to $\mu^2$ and scale with radius as $\propto r^{-11/2}$ \citep{lai99,pfe04}. Of these two effects, the magnetic torque seems to be the most
efficient one, as \citet{pfe04} state that $r_{\rm BP}\sim3r_{\rm in}\sim10R_*$, where $r_{\rm in}\propto\mu^{4/7}$ is the is the magnetospheric truncation
radius. On the other hand, for a rapidly spinning oblate star (say, rotating at a tenth of the breakup rate) the transition radius is smaller: $r_{\rm BP}\approx R_* (\tfrac{3}{2}J_2 \alpha h^{-2})^{1/2}\sim\mathcal{O}(R_*)$. Thus, young stars can, in principle, enforce the spin-orbit alignment of the surrounding disk out to a distance of a few $R_*$, while regions outside that radius would be subject to external torques or able to retain any primordial misalignments.}

\section{Summary}
In this work, we have studied the obliquity distribution of exoplanet host stars,
performing an estimation of the concentration parameter
$\kappa$, which serves as a measure of how narrow/wide the distribution of obliquities
is around/away from $\psi=0^\circ$. We have carried out this parameter estimation 
using a hierarchical Bayesian inference method (\citealp{hog10,for14}, MW14), which we have applied to two
publicly available datasets. The first one, the CKS \citep{pet17,joh17}, helps constrain $\kappa$
from measurements/inference of the LOS inclination of the stellar spin angle $I_*$. The second dataset
\citep[TEPCat][]{sou11}, provides measurements of the projected obliquity onto the plane of the sky
$\lambda$, which can also be used to estimate $\kappa$ (Section~\ref{sec:overview}). From these
datasets, we can conclude:
\begin{itemize}
\item[$\bullet$] The CKS dataset of $V\sin I_*$ and $R_*$ provides us with
 a target list ($N=257$) with LOS inclination angles
that are consistent with a concentration parameter of 
$\kappa_{\text{\tiny CKS}}=14.5^{+13.5}_{-6}$,
larger than previously measured for {\it Kepler} systems {(MW14 found $\kappa_{\text{\tiny MW}}=6.2^{+1.8}_{-1.6}$)}
 and consistent with
a mean obliquity of $\langle\psi\rangle=19^\circ$ and a standard deviation of  $\sigma_\psi=10^\circ$.
 \item[$\bullet$] The TEPCat dataset of $\lambda$ measurements ($N=118$) is consistent with
 $\kappa_{\text{\tiny TEPCAT}}=2.2^{+0.2}_{-0.6}$, meaning that the class of systems probed by
 RM observations (typically hot Jupiters)
 follow a {\it significantly} wider distribution of obliquities
 than those probed by the {\it Kepler} telescope. 
\end{itemize}

We have divided the CKS and TEPCat datasets into separate
bins according to stellar and planetary properties. With the CKS catalog, we have explored
stellar age, stellar metallicity, stellar multiplicity, planet multiplicity, planet orbital period
and planet radius.  With the TEPCat catalog, we have explored stellar effective temperature.
Our findings are:
\begin{itemize}
\item[$\bullet$] Planet multiplicity does not affect stellar obliquity at a statistically significant level
in {\it Kepler} systems, and the weak trend originally pointed out by MW14 has vanished.
This finding is in agreement the recent results of \citet{win17}.
Although it is still true that compact multi-planet systems have very
low obliquities \citep[e.g.][]{san12,alb13}, the converse is not necessarily true for single systems
of any period and planet radius.  Hot Jupiter systems,  which
do tend to lack nearby neighbors \citep{ste12}, also
 tend to have higher obliquities on average  \citep[e.g.,][]{heb08,win09a,tri10,alb13}, but this property does not extend to
single-transiting systems with small planets.
\item[$\bullet$] {We have looked for trends in $\kappa$ as a function of stellar properties 
within the CKS set, exploring the
dependence on $T_{\rm eff}$, stellar multiplicity, stellar age and stellar metallicity.
The only obliquity trend that rises to a substantial statistical level ($\gtrsim2.5{-}\sigma$) is that of metallicity.
None of the other three stellar variables affects the inferred value of $\kappa$ above a 2-$\sigma$ level.}
\item[$\bullet$] A now well established trend is that hot Jupiters hosts with
$T_{\rm eff}\gtrsim6000$~K tend to be more oblique. This trend is impossible to probe
with the CKS set (it lacks hot Jupiter systems with $T_{\rm eff}>6000$~K), but
it is easily seen in the TEPCat catalog. By using concentration parameter
inference on the TEPCat catalog, we have not only recovered a known temperature trend,
but we have placed a {\it quantitative} measure of how much more oblique hotter hosts are.
{We find, when considering hot Jupiters alone (106 objects) $\kappa^{\text{ \tiny TEPCAT,HJ}}_{\rm hotter}=0.9\pm0.3$ and $\kappa^{\text{ \tiny TEPCAT,HJ}}_{\rm cooler}=4.2^{+0.5}_{-0.8}$ with large statistical significance. Despite the greater obliquity of hot Jupiter host above
the Kraft break, both concentration parameters $\kappa^{\text{ \tiny TEPCAT,HJ}}_{\rm hotter}$ and $\kappa^{\text{ \tiny TEPCAT,HJ}}_{\rm cooler}$
 imply wide distributions of $\psi$}. Thus, {\it all hot Jupiters hosts} in the TEPCat
catalog are significantly more oblique than the overall {\it Kepler} systems, independently of the stellar effective temperature.
\item[$\bullet$] {We have looked for trends in $\kappa$ as a function of planetary properties 
within the CKS set.  We find a trends with planetary orbital period at a substantial level ($2.5{-}\sigma$) period 
and a trend with planetary radius at a convincing level (${\sim}3{-}\sigma$). 
 We propose the hypothesis
that the correlation between obliquity and planet radius is indirectly probing a un underlying ``dynamical temperature'' trend.  Under such hypothesis, 
the more tightly packed systems or those with higher total mass contained in planets will exhibit larger stellar obliquities. This is in accordance
with the trend found with stellar metallicity.
On the other hand, we propose that spin-orbit alignment in short-period systems is due to primordial alignment of the innermost protoplanetary gas disk, and not due
to tidal realignment of the host star at later times.}
\item[$\bullet$] 	We have attempted to study the obliquity for hot Jupiters in the {\it Kepler} catalog.
Given the small number of these objects for which all $V\sin I_*$, $R_*$ and $P_{\rm rot}$
can be compiled, it is difficult to provide a precise number. {However, the data favors isotropic orientations over spin-orbit alignment, which is in rough agreement with the TEPCat results.}
\end{itemize}

One of the remaining challenges when studying the statistical distribution of obliquity is
being able to  robustly measure the role of the Kraft break in stellar alignment. Although
the $V\sin I_*$ method is observationally cheaper and less biased toward a specific type of planet
than than RM observations, detecting small planets around stars with $T_{\rm eff}>6000$~K is
difficult, as is the unambiguous identification of rotational periods.  Of these two difficulties, perhaps 
the most concerning is {the frequent lack of reliable periods for the hotter stars with detected planets} \citep[see][]{maz15,win17}.
{The QPGP method of period identification seems promising when unveiling underlying rotational periods of active stars, although
the possibility of significant differential rotation} \citep[e.g.,][]{rei06}  {might require extensions to this parametric model.
The identification of stellar $\sin I_*$ from future asteroseismological studies can be instrumental in providing an independent confirmation of
the inference based on  the $V\sin I_*$ approach.}

\vspace{0.1in}

We have provided open-source, 
documented software on how to carry out the hierarchical Bayesian inference of the concentration parameter $\kappa$. This software package
has been made freely available online (\url{github.com/djmunoz/obliquity_inference}).

\vspace{-0.1in}
\acknowledgements{
We acknowledge support from the Israel Science Foundation I-CORE 1829/12 program, and the Minerva Center for 
Life Under Extreme Planetary Conditions.
DJM is grateful to Josh Winn for feedback and advice at an early stage of this project,
to Amaury Triad for constructively critical feedback on the manuscript,  to Sam Quinn for patiently 
explaining the limitations of rotational velocity measurements, and to Arieh K\"onigl for fruitful discussion.
We also thank Ruth Angus for generously sharing an updated database of rotational periods, Ben Nelson
Maxwell Moe and Crist\'obal Petrovich for enriching discussions, and Erik Petigura, Gongjie Li, Bekki Dawson and Gijs Mulders  for
comments on the manuscript.
DJM thanks all the members of Astrophysics Group at the Technion Physics Department,
with special appreciation to Evgeni Grishin, Erez Michaely, Serena Repetto and Yossef Zenati, for
stimulating discussions and for providing a welcoming environment during a rewarding 6-month visit in Haifa, Israel.}

\vspace{-0.1in}

\bibliographystyle{apj}

\begin{thebibliography}{}
\expandafter\ifx\csname natexlab\endcsname\relax\def\natexlab#1{#1}\fi

\bibitem[{{Aigrain} {et~al.}(2015){Aigrain}, {Llama}, {Ceillier}, {Chagas},
  {Davenport}, {Garc{\'{\i}}a}, {Hay}, {Lanza}, {McQuillan}, {Mazeh}, {de
  Medeiros}, {Nielsen}, \& {Reinhold}}]{aig15}
{Aigrain}, S., {Llama}, J., {Ceillier}, T., {et~al.} 2015, \mnras, 450, 3211

\bibitem[{{Albrecht} {et~al.}(2011){Albrecht}, {Winn}, {Carter}, {Snellen}, \&
  {de Mooij}}]{alb11}
{Albrecht}, S., {Winn}, J.~N., {Carter}, J.~A., {Snellen}, I.~A.~G., \& {de
  Mooij}, E.~J.~W. 2011, \apj, 726, 68

\bibitem[{{Albrecht} {et~al.}(2013){Albrecht}, {Winn}, {Marcy}, {Howard},
  {Isaacson}, \& {Johnson}}]{alb13}
{Albrecht}, S., {Winn}, J.~N., {Marcy}, G.~W., {et~al.} 2013, \apj, 771, 11

\bibitem[{{Albrecht} {et~al.}(2012){Albrecht}, {Winn}, {Johnson}, {Howard},
  {Marcy}, {Butler}, {Arriagada}, {Crane}, {Shectman}, {Thompson}, {Hirano},
  {Bakos}, \& {Hartman}}]{alb12}
{Albrecht}, S., {Winn}, J.~N., {Johnson}, J.~A., {et~al.} 2012, \apj, 757, 18

\bibitem[{{Anderson} {et~al.}(2016){Anderson}, {Storch}, \& {Lai}}]{and16}
{Anderson}, K.~R., {Storch}, N.~I., \& {Lai}, D. 2016, \mnras, 456, 3671

\bibitem[{{Angus} {et~al.}(2018){Angus}, {Morton}, {Aigrain}, {Foreman-Mackey},
  \& {Rajpaul}}]{ang18}
{Angus}, R., {Morton}, T., {Aigrain}, S., {Foreman-Mackey}, D., \& {Rajpaul},
  V. 2018, \mnras, 474, 2094

\bibitem[{{Bailey} {et~al.}(2016){Bailey}, {Batygin}, \& {Brown}}]{bai16}
{Bailey}, E., {Batygin}, K., \& {Brown}, M.~E. 2016, \aj, 152, 126

\bibitem[{{Bardeen} \& {Petterson}(1975)}]{bar75}
{Bardeen}, J.~M., \& {Petterson}, J.~A. 1975, \apjl, 195, L65

\bibitem[{{Beck} \& {Giles}(2005)}]{bec05}
{Beck}, J.~G., \& {Giles}, P. 2005, \apjl, 621, L153

\bibitem[{{Benomar} {et~al.}(2014){Benomar}, {Masuda}, {Shibahashi}, \&
  {Suto}}]{ben14}
{Benomar}, O., {Masuda}, K., {Shibahashi}, H., \& {Suto}, Y. 2014, \pasj, 66,
  94

\bibitem[{{Bonomo} \& {Lanza}(2012)}]{bon12}
{Bonomo}, A.~S., \& {Lanza}, A.~F. 2012, \aap, 547, A37

\bibitem[{{Borucki} \& {Summers}(1984)}]{bor84}
{Borucki}, W.~J., \& {Summers}, A.~L. 1984, \icarus, 58, 121

\bibitem[{{Bou{\'e}} \& {Fabrycky}(2014)}]{bou14b}
{Bou{\'e}}, G., \& {Fabrycky}, D.~C. 2014, \apj, 789, 111

\bibitem[{{Buchhave} {et~al.}(2012){Buchhave}, {Latham}, {Johansen},
  {Bizzarro}, {Torres}, {Rowe}, {Batalha}, {Borucki}, {Brugamyer}, {Caldwell},
  {Bryson}, {Ciardi}, {Cochran}, {Endl}, {Esquerdo}, {Ford}, {Geary},
  {Gilliland}, {Hansen}, {Isaacson}, {Laird}, {Lucas}, {Marcy}, {Morse},
  {Robertson}, {Shporer}, {Stefanik}, {Still}, \& {Quinn}}]{buc12}
{Buchhave}, L.~A., {Latham}, D.~W., {Johansen}, A., {et~al.} 2012, \nat, 486,
  375

\bibitem[{{Buzasi} {et~al.}(2016){Buzasi}, {Lezcano}, \& {Preston}}]{buz16}
{Buzasi}, D., {Lezcano}, A., \& {Preston}, H.~L. 2016, Journal of Space Weather
  and Space Climate, 6, A38

\bibitem[{{Campante} {et~al.}(2016){Campante}, {Lund}, {Kuszlewicz}, {Davies},
  {Chaplin}, {Albrecht}, {Winn}, {Bedding}, {Benomar}, {Bossini}, {Handberg},
  {Santos}, {Van Eylen}, {Basu}, {Christensen-Dalsgaard}, {Elsworth}, {Hekker},
  {Hirano}, {Huber}, {Karoff}, {Kjeldsen}, {Lundkvist}, {North}, {Silva
  Aguirre}, {Stello}, \& {White}}]{cam16}
{Campante}, T.~L., {Lund}, M.~N., {Kuszlewicz}, J.~S., {et~al.} 2016, \apj,
  819, 85

\bibitem[{{Carrington}(1863)}]{car1863}
{Carrington}, R.~C. 1863, {Observations of the Spots on the Sun from November
  9, 1853, to March 24, 1861, Made at Redhill (London: Williams \& Norgate)}

\bibitem[{{Chaplin} {et~al.}(2011){Chaplin}, {Bedding}, {Bonanno}, {Broomhall},
  {Garc{\'{\i}}a}, {Hekker}, {Huber}, {Verner}, {Basu}, {Elsworth}, {Houdek},
  {Mathur}, {Mosser}, {New}, {Stevens}, {Appourchaux}, {Karoff}, {Metcalfe},
  {Molenda-{\.Z}akowicz}, {Monteiro}, {Thompson}, {Christensen-Dalsgaard},
  {Gilliland}, {Kawaler}, {Kjeldsen}, {Ballot}, {Benomar}, {Corsaro},
  {Campante}, {Gaulme}, {Hale}, {Handberg}, {Jarvis}, {R{\'e}gulo}, {Roxburgh},
  {Salabert}, {Stello}, {Mullally}, {Li}, \& {Wohler}}]{cha11}
{Chaplin}, W.~J., {Bedding}, T.~R., {Bonanno}, A., {et~al.} 2011, \apjl, 732,
  L5

\bibitem[{Conneely \& Boehnke(2007)}]{con07}
Conneely, K.~N., \& Boehnke, M. 2007, American journal of human genetics, 81 6,
  1158

\bibitem[{{Dawson}(2014)}]{daw14}
{Dawson}, R.~I. 2014, \apjl, 790, L31

\bibitem[{{Dawson} \& {Johnson}(2018)}]{daw18}
{Dawson}, R.~I., \& {Johnson}, J.~A. 2018, ArXiv e-prints, arXiv:1801.06117

\bibitem[{{Doyle} {et~al.}(1984){Doyle}, {Wilcox}, \& {Lorre}}]{doy84}
{Doyle}, L.~R., {Wilcox}, T.~J., \& {Lorre}, J.~J. 1984, \apj, 287, 307

\bibitem[{{Fabrycky} \& {Tremaine}(2007)}]{fab07}
{Fabrycky}, D., \& {Tremaine}, S. 2007, \apj, 669, 1298

\bibitem[{{Fabrycky} \& {Winn}(2009)}]{fab09}
{Fabrycky}, D.~C., \& {Winn}, J.~N. 2009, \apj, 696, 1230

\bibitem[{{Fisher} {et~al.}(1993){Fisher}, {Lewis}, \& {Embleton}}]{fisher93}
{Fisher}, N.~I., {Lewis}, T., \& {Embleton}, B.~J.~J. 1993, {Statistical
  Analysis of Spherical Data} (Cambridge University Press)

\bibitem[{{Fisher}(1953)}]{fis53}
{Fisher}, R. 1953, Proceedings of the Royal Society of London Series A, 217,
  295

\bibitem[{{Foreman-Mackey} {et~al.}(2014){Foreman-Mackey}, {Hogg}, \&
  {Morton}}]{for14}
{Foreman-Mackey}, D., {Hogg}, D.~W., \& {Morton}, T.~D. 2014, \apj, 795, 64

\bibitem[{{Furlan} {et~al.}(2017){Furlan}, {Ciardi}, {Everett}, {Saylors},
  {Teske}, {Horch}, {Howell}, {van Belle}, {Hirsch}, {Gautier}, {Adams},
  {Barrado}, {Cartier}, {Dressing}, {Dupree}, {Gilliland}, {Lillo-Box},
  {Lucas}, \& {Wang}}]{fur17}
{Furlan}, E., {Ciardi}, D.~R., {Everett}, M.~E., {et~al.} 2017, \aj, 153, 71

\bibitem[{{Garc{\'{\i}}a} {et~al.}(2014){Garc{\'{\i}}a}, {Ceillier},
  {Salabert}, {Mathur}, {van Saders}, {Pinsonneault}, {Ballot}, {Beck},
  {Bloemen}, {Campante}, {Davies}, {do Nascimento}, {Mathis}, {Metcalfe},
  {Nielsen}, {Su{\'a}rez}, {Chaplin}, {Jim{\'e}nez}, \& {Karoff}}]{gar14}
{Garc{\'{\i}}a}, R.~A., {Ceillier}, T., {Salabert}, D., {et~al.} 2014, \aap,
  572, A34

\bibitem[{{Gaudi} \& {Winn}(2007)}]{gau07}
{Gaudi}, B.~S., \& {Winn}, J.~N. 2007, \apj, 655, 550

\bibitem[{{Gim{\'e}nez}(2006)}]{gim06}
{Gim{\'e}nez}, A. 2006, \apj, 650, 408

\bibitem[{{Gizon} \& {Solanki}(2003)}]{giz03}
{Gizon}, L., \& {Solanki}, S.~K. 2003, \apj, 589, 1009

\bibitem[{{H{\'e}brard} {et~al.}(2008){H{\'e}brard}, {Bouchy}, {Pont},
  {Loeillet}, {Rabus}, {Bonfils}, {Moutou}, {Boisse}, {Delfosse}, {Desort},
  {Eggenberger}, {Ehrenreich}, {Forveille}, {Lagrange}, {Lovis}, {Mayor},
  {Pepe}, {Perrier}, {Queloz}, {Santos}, {S{\'e}gransan}, {Udry}, \&
  {Vidal-Madjar}}]{heb08}
{H{\'e}brard}, G., {Bouchy}, F., {Pont}, F., {et~al.} 2008, \aap, 488, 763

\bibitem[{{H{\'e}brard} {et~al.}(2011){H{\'e}brard}, {Ehrenreich}, {Bouchy},
  {Delfosse}, {Moutou}, {Arnold}, {Boisse}, {Bonfils}, {D{\'{\i}}az},
  {Eggenberger}, {Forveille}, {Lagrange}, {Lovis}, {Pepe}, {Perrier}, {Queloz},
  {Santerne}, {Santos}, {S{\'e}gransan}, {Udry}, \& {Vidal-Madjar}}]{heb11}
{H{\'e}brard}, G., {Ehrenreich}, D., {Bouchy}, F., {et~al.} 2011, \aap, 527,
  L11

\bibitem[{{Hirano} {et~al.}(2012){Hirano}, {Sanchis-Ojeda}, {Takeda}, {Narita},
  {Winn}, {Taruya}, \& {Suto}}]{hir12}
{Hirano}, T., {Sanchis-Ojeda}, R., {Takeda}, Y., {et~al.} 2012, \apj, 756, 66

\bibitem[{{Hirano} {et~al.}(2014){Hirano}, {Sanchis-Ojeda}, {Takeda}, {Winn},
  {Narita}, \& {Takahashi}}]{hir14}
---. 2014, \apj, 783, 9

\bibitem[{{Hogg} {et~al.}(2010){Hogg}, {Myers}, \& {Bovy}}]{hog10}
{Hogg}, D.~W., {Myers}, A.~D., \& {Bovy}, J. 2010, \apj, 725, 2166

\bibitem[{Holm(1979)}]{hol79}
Holm, S. 1979, Scand. J. Statist., 6, 65

\bibitem[{{Jeffreys}(1961)}]{jeffreys61}
{Jeffreys}, H. 1961, Theory of Probability, 3rd edn. (Oxford)

\bibitem[{{Johnson} {et~al.}(2017){Johnson}, {Petigura}, {Fulton}, {Marcy},
  {Howard}, {Isaacson}, {Hebb}, {Cargile}, {Morton}, {Weiss}, {Winn}, {Rogers},
  {Sinukoff}, \& {Hirsch}}]{joh17}
{Johnson}, J.~A., {Petigura}, E.~A., {Fulton}, B.~J., {et~al.} 2017, \aj, 154,
  108

\bibitem[{{Kaib} {et~al.}(2011){Kaib}, {Raymond}, \& {Duncan}}]{kai11}
{Kaib}, N.~A., {Raymond}, S.~N., \& {Duncan}, M.~J. 2011, \apjl, 742, L24

\bibitem[{{Kraft}(1967)}]{kra67}
{Kraft}, R.~P. 1967, \apj, 150, 551

\bibitem[{{Kumar} \& {Pringle}(1985)}]{kum85}
{Kumar}, S., \& {Pringle}, J.~E. 1985, \mnras, 213, 435

\bibitem[{{Lai}(1999)}]{lai99}
{Lai}, D. 1999, \apj, 524, 1030

\bibitem[{{Lai}(2016)}]{lai16}
---. 2016, \aj, 152, 215

\bibitem[{{Lai} {et~al.}(2018){Lai}, {Anderson}, \& {Pu}}]{lai18}
{Lai}, D., {Anderson}, K.~R., \& {Pu}, B. 2018, \mnras, arXiv:1710.11140

\bibitem[{{Lai} \& {Pu}(2017)}]{lai17}
{Lai}, D., \& {Pu}, B. 2017, \aj, 153, 42

\bibitem[{{Latham} {et~al.}(2011){Latham}, {Rowe}, {Quinn}, {Batalha},
  {Borucki}, {Brown}, {Bryson}, {Buchhave}, {Caldwell}, {Carter},
  {Christiansen}, {Ciardi}, {Cochran}, {Dunham}, {Fabrycky}, {Ford}, {Gautier},
  {Gilliland}, {Holman}, {Howell}, {Ibrahim}, {Isaacson}, {Jenkins}, {Koch},
  {Lissauer}, {Marcy}, {Quintana}, {Ragozzine}, {Sasselov}, {Shporer},
  {Steffen}, {Welsh}, \& {Wohler}}]{lat11}
{Latham}, D.~W., {Rowe}, J.~F., {Quinn}, S.~N., {et~al.} 2011, \apjl, 732, L24

\bibitem[{{Li} \& {Winn}(2016)}]{li16}
{Li}, G., \& {Winn}, J.~N. 2016, \apj, 818, 5

\bibitem[{{Matsakos} \& {K{\"o}nigl}(2015)}]{mat15}
{Matsakos}, T., \& {K{\"o}nigl}, A. 2015, \apjl, 809, L20

\bibitem[{{Mazeh} {et~al.}(2015){Mazeh}, {Perets}, {McQuillan}, \&
  {Goldstein}}]{maz15}
{Mazeh}, T., {Perets}, H.~B., {McQuillan}, A., \& {Goldstein}, E.~S. 2015,
  \apj, 801, 3

\bibitem[{{McLaughlin}(1924)}]{mcl24}
{McLaughlin}, D.~B. 1924, \apj, 60, doi:10.1086/142826

\bibitem[{{McQuillan} {et~al.}(2013){McQuillan}, {Aigrain}, \&
  {Mazeh}}]{mcq13a}
{McQuillan}, A., {Aigrain}, S., \& {Mazeh}, T. 2013, \mnras, 432, 1203

\bibitem[{{McQuillan} {et~al.}(2014){McQuillan}, {Mazeh}, \& {Aigrain}}]{mcq14}
{McQuillan}, A., {Mazeh}, T., \& {Aigrain}, S. 2014, \apjs, 211, 24

\bibitem[{Miller(1981)}]{mil81}
Miller, R. 1981, Simultaneous statistical inference, Springer series in
  statistics (Springer-Verlag)

\bibitem[{{Montet} {et~al.}(2017){Montet}, {Tovar}, \&
  {Foreman-Mackey}}]{mon17}
{Montet}, B.~T., {Tovar}, G., \& {Foreman-Mackey}, D. 2017, \apj, 851, 116

\bibitem[{{Morton} \& {Johnson}(2011)}]{mor11}
{Morton}, T.~D., \& {Johnson}, J.~A. 2011, \apj, 729, 138

\bibitem[{{Morton} \& {Winn}(2014)}]{mor14}
{Morton}, T.~D., \& {Winn}, J.~N. 2014, \apj, 796, 47

\bibitem[{{Mulders} {et~al.}(2016){Mulders}, {Pascucci}, {Apai}, {Frasca}, \&
  {Molenda-{\.Z}akowicz}}]{mul16}
{Mulders}, G.~D., {Pascucci}, I., {Apai}, D., {Frasca}, A., \&
  {Molenda-{\.Z}akowicz}, J. 2016, \aj, 152, 187

\bibitem[{{Mullally} {et~al.}(2015){Mullally}, {Coughlin}, {Thompson}, {Rowe},
  {Burke}, {Latham}, {Batalha}, {Bryson}, {Christiansen}, {Henze}, {Ofir},
  {Quarles}, {Shporer}, {Van Eylen}, {Van Laerhoven}, {Shah}, {Wolfgang},
  {Chaplin}, {Xie}, {Akeson}, {Argabright}, {Bachtell}, {Barclay}, {Borucki},
  {Caldwell}, {Campbell}, {Catanzarite}, {Cochran}, {Duren}, {Fleming},
  {Fraquelli}, {Girouard}, {Haas}, {He{\l}miniak}, {Howell}, {Huber}, {Larson},
  {Gautier}, {Jenkins}, {Li}, {Lissauer}, {McArthur}, {Miller}, {Morris},
  {Patil-Sabale}, {Plavchan}, {Putnam}, {Quintana}, {Ramirez}, {Silva Aguirre},
  {Seader}, {Smith}, {Steffen}, {Stewart}, {Stober}, {Still}, {Tenenbaum},
  {Troeltzsch}, {Twicken}, \& {Zamudio}}]{mull15}
{Mullally}, F., {Coughlin}, J.~L., {Thompson}, S.~E., {et~al.} 2015, \apjs,
  217, 31

\bibitem[{{Nagasawa} {et~al.}(2008){Nagasawa}, {Ida}, \& {Bessho}}]{nag08}
{Nagasawa}, M., {Ida}, S., \& {Bessho}, T. 2008, \apj, 678, 498

\bibitem[{{Naoz}(2016)}]{nao16}
{Naoz}, S. 2016, \araa, 54, 441

\bibitem[{{Naoz} {et~al.}(2012){Naoz}, {Farr}, \& {Rasio}}]{nao12}
{Naoz}, S., {Farr}, W.~M., \& {Rasio}, F.~A. 2012, \apjl, 754, L36

\bibitem[{{Nutzman} {et~al.}(2011){Nutzman}, {Fabrycky}, \& {Fortney}}]{nut11}
{Nutzman}, P.~A., {Fabrycky}, D.~C., \& {Fortney}, J.~J. 2011, \apjl, 740, L10

\bibitem[{{Ohta} {et~al.}(2005){Ohta}, {Taruya}, \& {Suto}}]{oht05}
{Ohta}, Y., {Taruya}, A., \& {Suto}, Y. 2005, \apj, 622, 1118

\bibitem[{{Papaloizou} \& {Pringle}(1983)}]{pap83}
{Papaloizou}, J.~C.~B., \& {Pringle}, J.~E. 1983, \mnras, 202, 1181

\bibitem[{{Paz-Chinch{\'o}n} {et~al.}(2015){Paz-Chinch{\'o}n}, {Le{\~a}o},
  {Bravo}, {de Freitas}, {Ferreira Lopes}, {Alves}, {Catelan}, {Canto Martins},
  \& {De Medeiros}}]{paz15}
{Paz-Chinch{\'o}n}, F., {Le{\~a}o}, I.~C., {Bravo}, J.~P., {et~al.} 2015, \apj,
  803, 69

\bibitem[{{Petigura} {et~al.}(2017){Petigura}, {Howard}, {Marcy}, {Johnson},
  {Isaacson}, {Cargile}, {Hebb}, {Fulton}, {Weiss}, {Morton}, {Winn}, {Rogers},
  {Sinukoff}, {Hirsch}, \& {Crossfield}}]{pet17}
{Petigura}, E.~A., {Howard}, A.~W., {Marcy}, G.~W., {et~al.} 2017, \aj, 154,
  107

\bibitem[{{Petigura} {et~al.}(2018){Petigura}, {Marcy}, {Winn}, {Weiss},
  {Fulton}, {Howard}, {Sinukoff}, {Isaacson}, {Morton}, \& {Johnson}}]{pet18}
{Petigura}, E.~A., {Marcy}, G.~W., {Winn}, J.~N., {et~al.} 2018, \aj, 155, 89

\bibitem[{{Petrovich}(2015{\natexlab{a}})}]{pet15b}
{Petrovich}, C. 2015{\natexlab{a}}, \apj, 805, 75

\bibitem[{{Petrovich}(2015{\natexlab{b}})}]{pet15a}
---. 2015{\natexlab{b}}, \apj, 799, 27

\bibitem[{{Pfeiffer} \& {Lai}(2004)}]{pfe04}
{Pfeiffer}, H.~P., \& {Lai}, D. 2004, \apj, 604, 766

\bibitem[{{Pu} \& {Wu}(2015)}]{pu15}
{Pu}, B., \& {Wu}, Y. 2015, \apj, 807, 44

\bibitem[{{Queloz} {et~al.}(2000){Queloz}, {Eggenberger}, {Mayor}, {Perrier},
  {Beuzit}, {Naef}, {Sivan}, \& {Udry}}]{que00}
{Queloz}, D., {Eggenberger}, A., {Mayor}, M., {et~al.} 2000, \aap, 359, L13

\bibitem[{{Quinn} \& {White}(2016)}]{qui16}
{Quinn}, S.~N., \& {White}, R.~J. 2016, \apj, 833, 173

\bibitem[{{Reiners}(2006)}]{rei06}
{Reiners}, A. 2006, \aap, 446, 267

\bibitem[{{Rossiter}(1924)}]{ros24}
{Rossiter}, R.~A. 1924, \apj, 60, doi:10.1086/142825

\bibitem[{{Sanchis-Ojeda} {et~al.}(2012){Sanchis-Ojeda}, {Fabrycky}, {Winn},
  {Barclay}, {Clarke}, {Ford}, {Fortney}, {Geary}, {Holman}, {Howard},
  {Jenkins}, {Koch}, {Lissauer}, {Marcy}, {Mullally}, {Ragozzine}, {Seader},
  {Still}, \& {Thompson}}]{san12}
{Sanchis-Ojeda}, R., {Fabrycky}, D.~C., {Winn}, J.~N., {et~al.} 2012, \nat,
  487, 449

\bibitem[{{Sanchis-Ojeda} {et~al.}(2013){Sanchis-Ojeda}, {Winn}, {Marcy},
  {Howard}, {Isaacson}, {Johnson}, {Torres}, {Albrecht}, {Campante}, {Chaplin},
  {Davies}, {Lund}, {Carter}, {Dawson}, {Buchhave}, {Everett}, {Fischer},
  {Geary}, {Gilliland}, {Horch}, {Howell}, \& {Latham}}]{san13}
{Sanchis-Ojeda}, R., {Winn}, J.~N., {Marcy}, G.~W., {et~al.} 2013, \apj, 775,
  54

\bibitem[{{Schatzman}(1962)}]{sch62}
{Schatzman}, E. 1962, Annales d'Astrophysique, 25, 18

\bibitem[{{Schlaufman}(2010)}]{sch10}
{Schlaufman}, K.~C. 2010, \apj, 719, 602

\bibitem[{{Soderblom}(1985)}]{sod85}
{Soderblom}, D.~R. 1985, \pasp, 97, 57

\bibitem[{{Southworth}(2011)}]{sou11}
{Southworth}, J. 2011, \mnras, 417, 2166

\bibitem[{{Steffen} {et~al.}(2012){Steffen}, {Ragozzine}, {Fabrycky}, {Carter},
  {Ford}, {Holman}, {Rowe}, {Welsh}, {Borucki}, {Boss}, {Ciardi}, \&
  {Quinn}}]{ste12}
{Steffen}, J.~H., {Ragozzine}, D., {Fabrycky}, D.~C., {et~al.} 2012,
  Proceedings of the National Academy of Science, 109, 7982

\bibitem[{{Struve}(1930)}]{str30}
{Struve}, O. 1930, \apj, 72, doi:10.1086/143256

\bibitem[{{Tremaine}(2015)}]{tre15}
{Tremaine}, S. 2015, \apj, 807, 157

\bibitem[{{Tremaine} \& {Dong}(2012)}]{tre12}
{Tremaine}, S., \& {Dong}, S. 2012, \aj, 143, 94

\bibitem[{{Triaud}(2011)}]{tri11}
{Triaud}, A.~H.~M.~J. 2011, \aap, 534, L6

\bibitem[{{Triaud} {et~al.}(2010){Triaud}, {Collier Cameron}, {Queloz},
  {Anderson}, {Gillon}, {Hebb}, {Hellier}, {Loeillet}, {Maxted}, {Mayor},
  {Pepe}, {Pollacco}, {S{\'e}gransan}, {Smalley}, {Udry}, {West}, \&
  {Wheatley}}]{tri10}
{Triaud}, A.~H.~M.~J., {Collier Cameron}, A., {Queloz}, D., {et~al.} 2010,
  \aap, 524, A25

\bibitem[{{Weiss} \& {Marcy}(2014)}]{wei14}
{Weiss}, L.~M., \& {Marcy}, G.~W. 2014, \apjl, 783, L6

\bibitem[{{Winn} {et~al.}(2010){Winn}, {Fabrycky}, {Albrecht}, \&
  {Johnson}}]{win10}
{Winn}, J.~N., {Fabrycky}, D., {Albrecht}, S., \& {Johnson}, J.~A. 2010, \apjl,
  718, L145

\bibitem[{{Winn} \& {Fabrycky}(2015)}]{win15}
{Winn}, J.~N., \& {Fabrycky}, D.~C. 2015, \araa, 53, 409

\bibitem[{{Winn} {et~al.}(2009{\natexlab{a}}){Winn}, {Johnson}, {Albrecht},
  {Howard}, {Marcy}, {Crossfield}, \& {Holman}}]{win09b}
{Winn}, J.~N., {Johnson}, J.~A., {Albrecht}, S., {et~al.} 2009{\natexlab{a}},
  \apjl, 703, L99

\bibitem[{{Winn} {et~al.}(2007){Winn}, {Holman}, {Henry}, {Roussanova}, {Enya},
  {Yoshii}, {Shporer}, {Mazeh}, {Johnson}, {Narita}, \& {Suto}}]{win07}
{Winn}, J.~N., {Holman}, M.~J., {Henry}, G.~W., {et~al.} 2007, \aj, 133, 1828

\bibitem[{{Winn} {et~al.}(2009{\natexlab{b}}){Winn}, {Johnson}, {Fabrycky},
  {Howard}, {Marcy}, {Narita}, {Crossfield}, {Suto}, {Turner}, {Esquerdo}, \&
  {Holman}}]{win09a}
{Winn}, J.~N., {Johnson}, J.~A., {Fabrycky}, D., {et~al.} 2009{\natexlab{b}},
  \apj, 700, 302

\bibitem[{{Winn} {et~al.}(2017){Winn}, {Petigura}, {Morton}, {Weiss}, {Dai},
  {Schlaufman}, {Howard}, {Isaacson}, {Marcy}, {Justesen}, \&
  {Albrecht}}]{win17}
{Winn}, J.~N., {Petigura}, E.~A., {Morton}, T.~D., {et~al.} 2017, \aj, 154, 270

\bibitem[{{Xie} {et~al.}(2016){Xie}, {Dong}, {Zhu}, {Huber}, {Zheng}, {De Cat},
  {Fu}, {Liu}, {Luo}, {Wu}, {Zhang}, {Zhang}, {Zhou}, {Cao}, {Hou}, {Wang}, \&
  {Zhang}}]{xie16}
{Xie}, J.-W., {Dong}, S., {Zhu}, Z., {et~al.} 2016, Proceedings of the National
  Academy of Science, 113, 11431

\bibitem[{{Zhu} {et~al.}(2018){Zhu}, {Petrovich}, {Wu}, {Dong}, \&
  {Xie}}]{zhu18}
{Zhu}, W., {Petrovich}, C., {Wu}, Y., {Dong}, S., \& {Xie}, J. 2018, ArXiv
  e-prints, arXiv:1802.09526

\end{thebibliography}


\appendix

\section{Projected Obliquity }\label{app:appendix_a}
\begin{figure*}[ht!]
\centering
\includegraphics[width=0.52\textwidth]{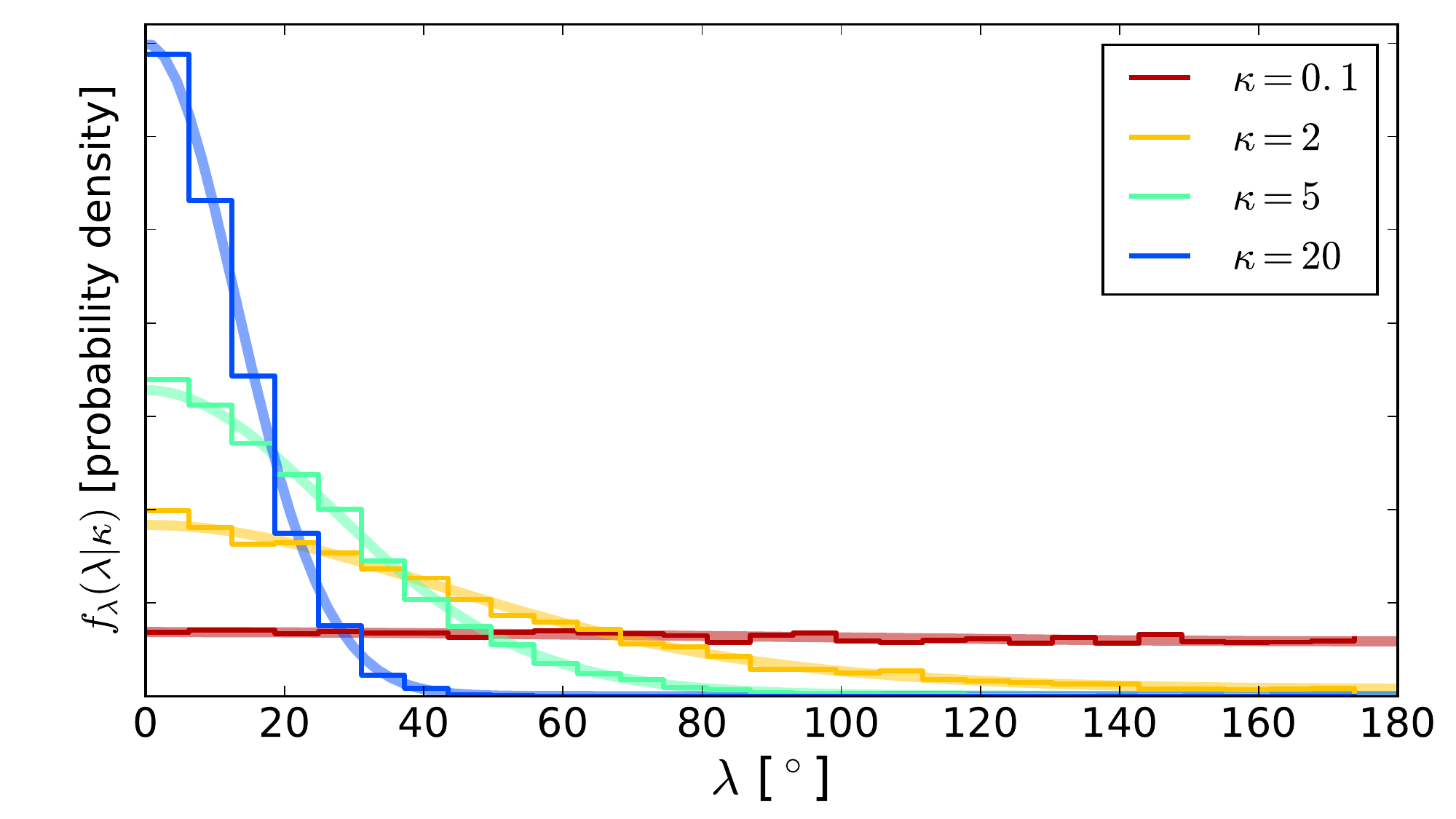}
\caption{PDF of $\lambda$ --the projected spin-orbit misalignment-- given difference values of $\kappa$
(the Fisher concentration parameter) as given by Eq.~(\ref{eq:lambda_pdf}). Histograms are constructed
after computing $\lambda$ from Monte Carlo samples ($N=2000$) of $\theta$ and $\phi$ 
(Eqs.~\ref{eq:angle_transform_b}-\ref{eq:angle_transform_c}), where $\theta$ is sampled
 from the Fisher distribution in Eq.~(\ref{eq:fisher_dist})
and $\phi\sim U(0,\pi/2)$~.
\label{fig:lambda_pdf}}
\end{figure*}

The three-dimensional orientation of the stellar spin vector can be written in terms of
the polar and azimuthal angles $\theta$ and $\phi$ \citep{fab09,mor14} where the polar axis
is $\hat{\mathbf{z}}$ -- the unit angular momentum vector of the planetary orbit -- or in terms
of the angles $I_*$ and $\lambda$ (LOS-inclination and projected obliquity), in which the LOS vector $\hat{\mathbf{x}}$
plays the role of the polar axis.  These two sets of angles are related by (see Eq.~\ref{eq:angle_transform})
\begin{subequations}\label{eq:angle_transforms}
\begin{align}
\cos I_* =\sin\theta\cos\phi~~,\\
\label{eq:angle_transform_b}
\sin I_* \sin\lambda=\sin\theta\sin\phi~~,\\
\label{eq:angle_transform_c}
\sin I_* \cos \lambda =\cos\theta~~.
\end{align}
\end{subequations}
Given the PDFs $f_\theta(\theta|\kappa)$ and $f_\phi(\phi)$, one can obtain the PDF $f_{\cos I_*}(\cdot|\kappa)$
using coordinate transformations \citep[][; their equation~9]{mor14}. In a similar fashion, since $\tan\lambda=\tan\theta\sin\phi$ 
(Eqs.~\ref{eq:angle_transform_b}-\ref{eq:angle_transform_c}), one can obtain a PDF for the quantity $\tan\lambda$ after obtaining the PDFs
$f_{\tan\theta}(\cdot |\kappa)$ and $f_{\sin\phi}(\cdot| \kappa)$:
\begin{gather}
\begin{align}
f_{\tan\theta}(y|\kappa)&=\frac{\kappa}{2\sinh\kappa}\frac{|y|}{(1+y^2)^{3/2}}
\exp\bigg({\frac{\kappa}{\sqrt{1+y^2}}}\frac{y}{|y|}\bigg),~~~y \in(-\infty,\infty)
\\
f_{\sin\phi}(x|\kappa)&=\frac{2}{\pi}\frac{1}{\sqrt{1-x^2}},\;\;\;\;x \in(0,1)
\end{align}
\end{gather}
with $y=\tan\theta$ and $x=\sin\phi$. Combining these two expressions and using that
\begin{equation}
f_{\tan\lambda}(z|\kappa)=\int_{-\infty}^{\infty}f_{\tan\theta}(y|\kappa)f_{\sin\phi}(z/y~|\kappa) \frac{1}{|y|}dy~~,
\end{equation}
we have, for $z=\tan\lambda$,
\begin{equation}
\begin{split}
f_{\tan\lambda}(z|\kappa)&=\frac{\kappa}{\pi\sinh\kappa}{\times}\left\{
\begin{array}{lc}
\displaystyle\int\limits_{z}^{\infty}
 dy\frac{\exp({{\kappa}/{\sqrt{1+y^2}}})}{(1+y^2)^{3/2}\sqrt{1-(z/y)^2}}& z>0\\
\displaystyle\int\limits_{|z|}^{\infty}
 dy\frac{\exp(-{{\kappa}/{\sqrt{1+y^2}}})}{(1+y^2)^{3/2}\sqrt{1-(z/y)^2}}&z<0
 \end{array}\right.{=}\frac{\kappa}{\pi\sinh\kappa}\int\limits_{0}^{1/\sqrt{1+z^2}}
d\tilde{y}\frac{\exp(\kappa\tilde{y}\,z/|z|)}{\sqrt{1-\tilde{y}^2}}\frac{\tilde{y}}{\sqrt{1-\frac{\tilde{y}^2}{1-\tilde{y}^2}z^2}}
\end{split}
\end{equation}
This expression is valid in the range of $(0,\infty)$, and thus not
very practical for random sampling purposes.
However, using that $f_{\tan \lambda}|d\tan\lambda /d\lambda|= f_\lambda$, we can write:
\begin{equation}\label{eq:lambda_pdf}
\begin{split}
f_{\lambda}(\lambda|\kappa)&=\frac{\kappa/\sinh\kappa}{\pi\cos^2\lambda}\int\limits_{0}^{\cos\lambda}
 d\tilde{y}\frac{\exp({{\kappa}\tilde{y}})}{\sqrt{1-\tilde{y}^2}}\frac{\tilde{y}}{\sqrt{1-\tan^2\lambda\tfrac{\tilde{y}^2}{1-\tilde{y}^2}}}
\end{split}
\end{equation}
which we can confirm is correct by random sampling in $\theta$ and $\phi$ and computing the sampled values
of $\lambda$ using Eq.~\ref{eq:angle_transforms} (see Fig.~\ref{fig:lambda_pdf}).

\section{Background: Hierarchical Bayesian Inference of Meta-parameters}\label{app:background}
Here, we briefly describe the hierarchical Bayesian method introduced by \citet{hog10}. In this method, one seeks  to carry out the estimation
of the parameter vector ${\boldsymbol\alpha}$, which controls the probability distribution function (PDF) $f_{\boldsymbol\omega}(\boldsymbol\omega|{\boldsymbol\alpha})$
of a vector of physical quantities $\boldsymbol\omega$,
which in turn contains the parameters $\{\omega_i\}$ of some parametrization of an individual object.
With $N$ of such objects (in our case, exoplanet systems) the whole dataset $D$ can be split into subsets $D_1,...,D_N$, with $D=\cup_{n=1}^N D_n$.
By means of Bayes' theorem, the posterior PDF of these ``meta-parameters" ${\boldsymbol\alpha}$ for a given dataset $D$ is 
\citep{hog10}
\begin{equation}\label{eq:posterior_pdf}
\begin{split}
p({\boldsymbol\alpha}|D)&\propto
\mathcal{L}_{{\boldsymbol\alpha}}(\{D_n\}_{n=1}^N | {\boldsymbol\alpha}) \,\pi_{\boldsymbol\alpha}({\boldsymbol\alpha})\\
&=\left[\prod_{n=1}^Np(D_n| {\boldsymbol\alpha}) \right] \,\pi_{\boldsymbol\alpha}({\boldsymbol\alpha})\\
&\equiv\left[\prod_{n=1}^N\mathcal{L}_{{\boldsymbol\alpha},n} \right] \,\pi_{\boldsymbol\alpha}({\boldsymbol\alpha})~~,
\end{split}
\end{equation}
where $\mathcal{L}_{{\boldsymbol\alpha}}$ is the global likelihood, $\mathcal{L}_{{\boldsymbol\alpha}}$ is the likelihood associated to dataset
$D_n$ and $\pi_{\boldsymbol\alpha}$ the prior information of the parameter vector ${\boldsymbol\alpha}$.

Instead of computing $\mathcal{L}_{{\boldsymbol\alpha},n}(D_n | {\boldsymbol\alpha})$ directly from the global dataset, 
this hierarchical method introduces a middle step, which is played by the vector of parameters {\it per object} ${\boldsymbol\omega}_n$, 
for which we already have previously computed posterior PDF obtained from its corresponding dataset $D_n$:
\begin{equation}
p({\boldsymbol\omega}_n |D_n) = \frac{1}{Z_n}p(D_n|{\boldsymbol\omega}_n)\pi_{{\boldsymbol\omega},0}({\boldsymbol\omega}_n)
\end{equation}
where $Z_n$ is a normalization constant and $\pi_{{\boldsymbol\omega},0}$ is the (uninformative) prior for the vector of parameters ${\boldsymbol\omega}$.
Then, one can write \citep{hog10,for14}
\begin{equation}\label{eq:delta_likelihood}
\begin{split}
\mathcal{L}_{{\boldsymbol\alpha},n} & =\int d{\boldsymbol\omega}_n p(D_n| {\boldsymbol\omega}_n) p ({\boldsymbol\omega}_n | {\boldsymbol\alpha}) \\
&=Z_n \int d{\boldsymbol\omega}_n \frac{p({\boldsymbol\omega}_n |D_n) }{\pi_{{\boldsymbol\omega},0}({\boldsymbol\omega}_n)}p ({\boldsymbol\omega}_n | {\boldsymbol\alpha})
\end{split}
\end{equation}
The PDFs in Equations~(\ref{eq:cosi_given_kappa}) and ~(\ref{eq:lambda_given_kappa}) can replace
$p ({\boldsymbol\omega}_n | {\boldsymbol\alpha})$ in
Eq.~(\ref{eq:delta_likelihood}) --with the parameter vectors ${\boldsymbol\omega}_n$ and ${\boldsymbol\alpha}$ having only one element each --  
to carry out the hierarchical inference calculation on $\kappa$.

Ignoring the normalization constant, $\mathcal{L}_{{\boldsymbol\alpha},n}$ can be interpreted as the distribution-weighted average of
the ratio $p ({\boldsymbol\omega}_n | {\boldsymbol\alpha})/\pi_{{\boldsymbol\omega},0}({\boldsymbol\omega}_n)$, which allows for a Monte-Carlo
approximation via $K-$sampling \citep{hog10}
\begin{equation}\label{eq:delta_likelihood_ksamp}
\mathcal{L}_{{\boldsymbol\alpha},n} = \left\langle
\frac{p ({\boldsymbol\omega}_n | {\boldsymbol\alpha})}{\pi_{{\boldsymbol\omega},0}({\boldsymbol\omega}_n)}
\right\rangle_{{{\boldsymbol\omega}_n}}\approx \frac{1}{K}\sum\limits_{k=1}^K
\frac{p ({\boldsymbol\omega}_{n,k} | {\boldsymbol\alpha})}{\pi_{{\boldsymbol\omega},0}({\boldsymbol\omega}_{n,k})}
\end{equation}
where ${\boldsymbol\omega}_{n,k}$ is the $k$-th random sample of ${\boldsymbol\omega}_{n}$ obtained from the posterior $p({\boldsymbol\omega}_n |D_n)$.
Although very powerful, the $K$-sampling approximation is only justified for high-dimension integrals \citep[as in the 7-variable case of][]{hog10}. For the one-dimensional
integrals of interest here (see below), $K$-sampling is not necessary and we instead use direct numerical integration.

 %


\end{document}